\documentclass[final]{elsarticle}

\makeatletter
\let\ps@pprintTitle\ps@plain
\makeatother

\usepackage[hidelinks,hyperfootnotes=false, bookmarks]{hyperref}
\usepackage{graphicx} 
\usepackage{mathtools}
\usepackage{amsmath}
\usepackage{amssymb}
\usepackage{amsthm}
\usepackage{bm}
\usepackage{placeins}
\usepackage{cases}
\usepackage{algorithm}
\usepackage{empheq}
\usepackage{algpseudocode}
\usepackage[utf8]{inputenc}
\usepackage{multirow}
\usepackage{booktabs}
\usepackage{colortbl}
\usepackage{color,soul}
\usepackage{array}
\biboptions{sort&compress}
\usepackage{nicefrac}
\usepackage{import}
\usepackage{epstopdf}
\usepackage{mathrsfs}
\usepackage{mathtools}
\usepackage[figurename=Fig.]{caption}
\usepackage[numbers]{natbib}
\usepackage{array}
\usepackage{amsfonts}
\usepackage{scalerel}
\usepackage[labelformat=simple]{subcaption}
\usepackage[scientific-notation=true]{siunitx}

\usepackage{blindtext, xcolor}
\usepackage{comment}
\usepackage{svg}
\usepackage{url}
\svgpath{{ps/}}

\usepackage{tikz}
\usepackage{amstext}
\usetikzlibrary{calc,arrows,shapes,positioning,backgrounds,fit}

\usepackage{svg}
\svgpath{{ps/}}

\DeclareMathOperator{\tr}{tr}

\newcolumntype{M}[1]{>{\centering\arraybackslash}m{#1}}

\newcolumntype{L}{>{\centering\arraybackslash}m{3cm}}
\usepackage[margin=1in]{geometry}
\usepackage{lineno,hyperref}

\algblock{Input}{EndInput}
\algnotext{EndInput}
\algblock{Output}{EndOutput}

\definecolor{greydark}{RGB}{120, 120, 120}
\definecolor{greylight}{RGB}{190, 190, 190}
\DeclareRobustCommand\full  {\tikz[baseline=-0.6ex]\draw[very thick] (0,0)--(0.5,0);}
\DeclareRobustCommand\dotted{\tikz[baseline=-0.6ex]\draw[ultra thick,dotted] (0,0)--(0.46,0);}
\DeclareRobustCommand\dashed{\tikz[baseline=-0.6ex]\draw[very thick,dashed] (0,0)--(0.54,0);}

\newcommand{\Tau}{\mathrm{T}}
\newcommand{\Bu}{\bm{u}}
\newcommand{\Be}{\bm{\varepsilon}}

\newcommand{\Ber}{\bm{\varepsilon}^\mathrm{r}}
\newcommand{\Bep}{\bm{\varepsilon}^\mathrm{p}}

\newcommand{\Bsig}{\bm{\sigma}}

\newcommand{\Bsp}{\bm{s}^\mathrm{p}}

\newcommand{\Bsr}{\bm{s}^\mathrm{r}}

\newcommand{\squad}{\hspace{0.5em}}
\newcommand{\bu}{\boldsymbol{u}}

\newcommand{\Bvep}{\bm{\varepsilon}^\mathrm{p}}

\newcommand{\Bvu}{\bm{u}}
\newcommand{\Bvsig}{\bm{\sigma}}

\usepackage[color=gray!20, backgroundcolor=yellow, textwidth=2cm, textsize=footnotesize]{todonotes}

\definecolor{myred}{RGB}{163,0,0}
\definecolor{myblue}{RGB}{0,67,104}
\definecolor{mygreen}{RGB}{0,104,0}

\usepackage{setspace}
\begin{document}

\begin{frontmatter}

\title{Phase-field modeling of cyclic behavior in quasi-brittle materials: A micromechanics-based approach}

\author[add1]{Mina Sarem}
\cortext[MS]{Corresponding author: Mina Sarem}
\ead{mina.sarem@kuleuven.be}
\author[add2]{Nuhamin Eshetu Deresse}
\ead{nuhamineshetu2@gmail.com}
\author[add2]{Els Verstrynge}
\ead{els.verstrynge@kuleuven.be}
\author[add1]{Stijn François}
\ead{stijn.francois@kuleuven.be}

\address[add1]{KU Leuven, Department of Civil Engineering, Structural Mechanics Section, Kasteelpark Arenberg 40, B-3001 Leuven, Belgium}
\address[add2]{KU Leuven, Department of Civil Engineering, Materials and Constructions Section, Kasteelpark Arenberg 40, B-3001 Leuven, Belgium}

\tnotetext[t1]{{\itshape Postprint version.}}
\tnotetext[t2]{{\itshape Published version:}
        M. Sarem, N. E. Deresse, E. Verstrynge and S.~Fran\c{c}ois.
       \newblock Phase-field modeling of cyclic behavior in quasi-brittle materials: A micromechanics-based approach.
       \newblock {\em European Journal of Mechanics - A/Solids}, 106263, 2026.\\
       {\itshape DOI: \tt\url{https://doi.org/10.1016/j.euromechsol.2026.106263}} \vspace{3mm}}
\begin{abstract}
In this paper, we extend a micromechanics-based phase-field framework for fatigue fracture to incorporate cyclic plasticity with ratcheting. This mechanism is particularly relevant for low-cycle fatigue, where the accumulation of inelastic strains plays a critical role in the progression to final failure. An energetic formulation is proposed in which the ratcheting strain is explicitly incorporated into both the free energy and the dissipation potential. Ratcheting is modeled within a pressure-dependent, non-associative plasticity framework through the evolution of a ratcheting strain that progressively accumulates over loading cycles, capturing the characteristic inelastic strain growth of cyclic plasticity in a thermodynamically consistent manner. The plastic potential is formulated such that the deviatoric and volumetric components of ratcheting can be controlled independently. A staggered solution scheme is employed to solve for the internal variables, including the ratcheting strain. The model is validated through numerical simulations under a wide range of loading conditions, including monotonic and cyclic regimes (stress- and strain-controlled, low- and high-cycle fatigue), enabling evaluation of its performance and the influence of ratcheting on the material response, and demonstrating its applicability to realistic engineering scenarios.


\end{abstract}
\end{frontmatter}

\clearpage

\section*{Nomenclature}

\begin{table}[h!]
\begin{tabular}{ll}
\hline
$\phi$					& Dissipation potential \\
$\psi$					& Free energy density \\
$\bm{\mathsf{C}}^\mathrm{dam}$		& $\alpha$-dependent damaged elasticity tensor \\
$h$					& Fatigue degradation function \\
$f^\mathrm{d}$		& Damage yield function \\
$f^\mathrm{p}$		& Plastic yield function \\
$g$					& Degradation function \\
$g^\mathrm{p}$ 		& Plastic potential \\
$\bm{\mathsf{H}}^\mathrm{kin}$		& $\alpha$-dependent kinematic hardening tensor \\
$\bm{\mathsf{I}}$	& Fourth-order symmetric identity tensor \\
$\hat{n}$				& Direction of the plastic strain tensor $\Bep$ \\
$\hat{\bm{r}}$		& Direction of the ratcheting strain tensor $\Ber$ \\
$s^\mathrm{d}$		& Thermodynamic conjugate of $\alpha$ \\
$\Bsp$				& Thermodynamic conjugate of $\Bep$ \\
$\bm{s}^\mathrm{r}$	& Thermodynamic conjugate of $\Ber$ \\
$\bar{\boldsymbol{t}}$		& Traction on $\partial\Omega_\mathrm{N}$ \\
$\bu$						& Displacement field \\ 
$\bar{\bu}$					& Displacement on $\partial\Omega_\mathrm{D}$ \\
$\alpha$					& Crack phase-field \\
$\mathcal{F}$					& Fatigue variable \\
$\vartheta$						& Fatigue history variable \\
$\partial\Omega_\mathrm{D}$		& Dirichlet boundary \\
$\partial\Omega_\mathrm{N}$		& Neumann boundary \\
$\Bep$				& Plastic strain tensor \\
$\Ber$				& Ratcheting strain tensor \\ 
$A_\theta$							& Constant dilation coefficient \\
$A_\varphi$				& Constant friction coefficient \\
$b$						& Degradation constant \\
$b_K$					& Bulk homogenization constant\\
$b_\mu$					& Shear homogenization constant\\
$E$						& Young's modulus \\
$G_\mathrm{cI}$      	& Mode I fracture energy \\
$G_\mathrm{cII}$      	& Mode II fracture energy \\
$k$						& Fatigue degradation parameter \\
$K$						& Bulk modulus \\
$\ell$					& Damage characteristic length \\
$\beta$					& Ratcheting constant \\
$\mathcal{F}_0$		& Fatigue threshold \\
$\theta$				    & Dilation angle \\
$\mu$					& Shear modulus \\
$\nu$					& Poisson's ratio \\
$\varphi$				& Friction angle \\ 
$s$                   & Smoothstep function \\
$\epsilon$           & Regularization length controlling the width of the smooth transition zone \\
$\xi$                   & Normalized smoothing parameter \\
$\lambda$			& Lagrange (plastic) multiplier \\
$\gamma$						& Accumulated plastic multiplier  \\ 
\hline
\end{tabular}
\end{table}

\section{Introduction}\label{intro}

Failure occurs when a structure can no longer safely sustain applied loads or perform its intended function. Among the various mechanisms that lead to such failure, cyclic loading is one of the most critical and widespread \cite{schijve2009a}. The resulting degradation process, generally known as fatigue, progressively weakens the material and can lead to failure at stress levels well below its monotonic load-bearing capacity \cite{christensen2013a}. In quasi-brittle or pressure-sensitive materials such as concrete, masonry and rock, cyclic loading may also induce ratcheting, a progressive accumulation of plastic strain under asymmetric cycles, which further contributes to long-term deterioration. Understanding the mechanisms behind cyclic failure is therefore essential for designing structures that remain safe and functional throughout their service life.

Experiments are a crucial first step in studying material behavior, and research has produced increasingly advanced testing techniques \cite{silva2023a}. Experimental studies under cyclic loading have investigated fatigue phenomena using techniques such as acoustic emission and digital image correlation \cite{deresse2022a, deresse2023a, deresse2024a}. While experiments help refine constitutive models by providing insights into essential parameters, they are often time-consuming and costly. Constitutive modeling thus offers a promising way to complement and reduce the need for extensive testing, though finding a balance between accuracy and efficiency remains a critical challenge.

In constitutive modeling, material behavior is typically described using macroscale mathematical models that capture key phenomena such as damage and plasticity through internal variables \cite{simo1998a}. While macroscale phenomenological models are widely used in engineering for their adaptability, they often rely on parameters that require extensive calibration and offer limited insight into the underlying failure mechanisms. As an alternative, micromechanical models provide physics-based insights that are crucial for accurately representing the microstructure of quasi-brittle materials \cite{you2024a}. The presence of microcracks results in distinct failure modes: brittle failure under tensile loading and shear failure governed by frictional sliding of closed microcracks under compressive loading \cite{choo2018a}, highlighting the importance of incorporating such microscale mechanisms into numerical models.

\begin{figure}[htb!]
  \centering
    \includegraphics[scale=1.2]{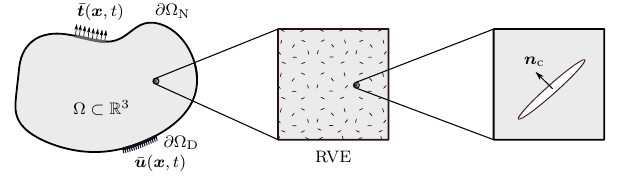}
\caption{Schematic representation of the continuum solid (left), the RVE consisting of a matrix-inclusion system with microcracks (middle), and the orientation of a penny-shaped microcrack (right) (adapted from Ulloa et al. \cite{ulloa2022a}).}
\label{fig1:micro}
\end{figure}

To address this, micromechanics-based phase-field modeling has emerged as a physically motivated alternative \cite{you2021b, ulloa2022a, wang2023a}. As illustrated in figure \ref{fig1:micro}, a zoomed-in view of a point in the solid shows the representative volume element (RVE), which captures the material's behavior at a smaller scale. Within the RVE, the model assumes that penny-shaped microcracks are randomly distributed throughout the solid matrix, and are considered either open or closed at any given time \cite{zhu2008b, ulloa2022a}. The effect of these microcracks at the microscale is linked to the macroscale response using homogenization techniques, such as the Mori–Tanaka scheme \cite{mori1973a}, which averages local fields to determine the macroscopic material behavior \cite{zhu2008d}. In this framework, microcrack opening is associated with damage (mode I fracture) and sliding corresponds to plasticity (mode II fracture) \cite{zhu2008a, liu2021b, zhao2018b}, naturally capturing tensile and shear failure modes without relying on heuristic energy splits. One of the key strengths of this framework is its ability to capture non-associative behavior characteristic of quasi-brittle materials through a Drucker–Prager formulation.

Fatigue phase-field models can be broadly categorized into those that introduce fatigue through an additional energetic contribution increasing the crack driving force \cite{schreiber2020a, haveroth2020a, loew2020a}, and those that degrade the fracture energy via a fatigue history variable \cite{alessi2018b, carrara2020a, seiler2020a, kristensen2020a, grossman2022a}. While effective at capturing high-cycle brittle fracture, these models do not account for inelastic strain accumulation, limiting their applicability to low-cycle fatigue scenarios where plasticity plays a decisive role. Ratcheting-driven cyclic plasticity within a pressure-dependent Drucker–Prager framework was introduced in variational form by Ulloa et al. \cite{ulloa2021b}. However, that work addressed plasticity only, without coupling to damage or fatigue. The subsequent work by Ulloa et al. \cite{ulloa2021a} coupled ratcheting to phase-field fatigue but within a classical phenomenological framework based on J2 plasticity, without pressure-dependence and without a micromechanical basis. The present work addresses these limitations by embedding ratcheting-driven cyclic plasticity within a micromechanics-based phase-field framework, where the constitutive equations are physically grounded in microcrack behavior and the coupling between damage, fatigue and pressure-dependent ratcheting emerges from a thermodynamically consistent formulation.

Ratcheting in quasi-brittle materials originates from microstructural changes such as compaction, pore collapse, and irreversible microcrack rearrangements \cite{liu2022a}, rather than the dislocation-based mechanisms typical of metals \cite{chaboche2008a}. Existing ratcheting models have been developed primarily for metals and are not suited to capture this distinct physical origin. Modeling ratcheting requires non-associative plasticity with an internal variable tracking the accumulated inelastic strain \cite{houlsby2017a}, and its coupling to fatigue damage in a thermodynamically consistent micromechanics-based framework has not been previously addressed.

The present paper fills this gap by extending the micromechanics-based variational phase-field framework for fatigue fracture \cite{sarem2025a} to incorporate ratcheting-driven cyclic plasticity. The key novelties are: (i) a thermodynamically consistent energetic formulation in which a ratcheting strain tensor is introduced as an independent internal variable entering both the free energy and the dissipation potential; (ii) a pressure-dependent, non-associative plastic potential with independent control of deviatoric and volumetric ratcheting components; and (iii) a staggered numerical implementation capable of capturing the coupled evolution of fracture and ratcheting-induced cyclic plasticity, demonstrated under monotonic, low-cycle, and high-cycle loading conditions; and (iv) validation of the proposed framework through comparison with experimental results under cyclic loading conditions.

This paper is organized as follows. section \ref{ratch} introduces the concept of ratcheting through a 1D schematic illustration. section \ref{ch6:ext} presents the formulation of the proposed model. The numerical implementation is outlined in section \ref{ch6:numimp}, followed by numerical simulations in section \ref{ch6:numsim}. Finally, section \ref{ch6:conc} summarizes the main conclusions.

\section{Cyclic plasticity via ratcheting}\label{ratch}


In quasi-brittle materials, inelastic deformation under cyclic loading does not necessarily follow the framework of classical plasticity, which is based on yield-surface evolution and dislocation motion as in metals. Instead, it can originate from microstructural changes such as local densification and pore collapse under compression \cite{liu2022a}. These mechanisms become increasingly active during cyclic loading, where the material gradually compacts over repeated cycles. These observations suggest that ratcheting in quasi-brittle materials is primarily driven by internal structural changes, such as compaction and microcrack evolution, rather than by the dislocation-based mechanisms typically observed in metals. Consequently, incorporating a formulation that captures cyclic plasticity with ratcheting effects is essential for reproducing the cyclic behavior of such materials, especially under compressive or non-proportional loading paths.

Ratcheting is characterized by the progressive accumulation of inelastic strain with each loading cycle \cite{alonso2004a}, and its manifestation depends on both the applied loading path and the material’s constitutive behavior. To model this behavior, the formulation proposed by Houlsby et al. \cite{houlsby2017a} is adopted, originally developed to describe the ratcheting behavior of cyclically loaded pile foundations \cite{abadie2015a}. The model is built on the classical framework of Armstrong and Frederick \cite{armstrong1966a}.

\begin{figure}[htb!]
 \vspace*{-1em}
  \centering
    \includegraphics[scale=1.2]{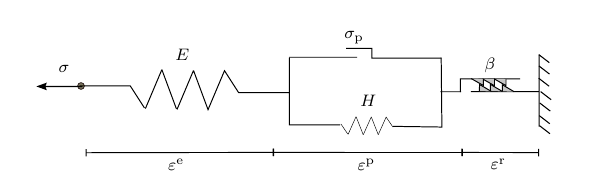}
\caption{1D rheological representation of elastoplastic-ratcheting model}
\label{fig6:ratch}
\end{figure}

To derive the governing equations, Houlsby et al. \cite{houlsby2017a} relied on the theory of hyperplasticity \cite{houlsby2007a}, which is closely related to the broader class of generalized standard materials and ensures thermodynamic consistency. Figure \ref{fig6:ratch} depicts the 1D rheological representation of an elastoplastic-ratcheting model, where ratcheting is in series with the plasticity and represents an additional mechanism of cyclic accumulation driven by plastic deformation. This schematic representation shows that whenever plastic strain occurs, an additional ratcheting strain $\Ber$ occurs which is a small fraction $\beta$ of the plastic strain. However, irrespective of the direction of the plastic strain, the ratcheting strain always occurs in the direction of the applied stress.

It is important to note that, although the evolution of the ratcheting strain is expressed in terms of the plastic strain increment, this does not imply that $\Ber$ is an additive component of the plastic strain. Instead, plastic strain acts as the driving mechanism for ratcheting, while $\Ber$ represents a distinct physical process associated with the gradual evolution of the material’s internal structure under cyclic loading. As will be shown in the thermodynamic formulation, $\Ber$ is treated as an independent internal variable with its own conjugate force and evolution law.

Following this definition, the ratcheting evolution is obtained as
\begin{equation}
\dot{\varepsilon}^{\mathrm{r}}=\mathrm{sign}(\sigma)\,\beta\,\Vert\dot{\varepsilon}^{\mathrm{p}}\Vert,
\label{c6:ratch}
\end{equation}
where $\Ber$ denotes the ratcheting strain, and the ratcheting parameter $\beta \in [0\, , \, 1]$ is a material constant that determines the fraction of plastic strains that contributes to ratcheting.

\section{Incorporation of ratcheting-driven cyclic plasticity}\label{ch6:ext}
A micromechanics-based phase-field model is proposed to capture both ratcheting-driven cyclic plasticity and fatigue. The formulation builds upon the energetic framework of the micromechanics-based phase-field model introduced in \cite{you2021b, ulloa2022a} and subsequently extended to fatigue in our earlier work \cite{sarem2025a}. Within this framework, damage dissipation at the macroscopic scale is described by the phase-field approach, while the evolution of plastic strain is governed by Drucker–Prager plasticity. The present section details the extension of the model to account for ratcheting effects through the incorporation of ratcheting strain into both the free energy and the dissipation potential in a thermodynamically consistent manner.

\subsection{Free energy}\label{ch3:free_ene}
The state of the solid in this problem is defined by three field variables: the displacement field $\Bu\colon\Omega\times\Tau\to\mathbb{R}^3$, the crack phase-field $\alpha\colon\Omega\times\Tau\to[0,1]$ to integrate the non-local effects into the local damage variable and the plastic strain tensor $\Bep\colon\Omega\times\Tau\to\mathbb{R}^{3\times 3}_\mathrm{sym}$ to incorporate a frictional sliding mechanism, and the ratcheting strain tensor $\Ber\colon\Omega\times\Tau\to\mathbb{R}^{3\times 3}_\mathrm{sym}$ to include irreversible ratcheting effects.

In the classical phase-field approach, the degradation function multiplies the elasticity tensor to reduce the energy during damage \cite{kuhn2015a, wu2020a}. However, this is not realistic as it results in symmetric behavior under tension and compression. To address this limitation, various energy decomposition techniques have been introduced in the literature \cite{freddi2011a,hakimzadeh2022a,vicentini2024a}, with the two main ones, spectral \cite{miehe2010a} and volumetric-deviatoric \cite{amor2009a} splits. In contrast, the micromechanics-based free energy formulation naturally captures asymmetric behavior without requiring an explicit energy split. This is achieved by distinguishing between open and closed microcracks. Based on this concept, the free energy density is expressed as:
\begin{equation}
\psi(\bm{\varepsilon},\bm{\varepsilon}^\mathrm{p},\bm{\varepsilon}^\mathrm{r},\alpha)\coloneqq\begin{dcases}
\frac{1}{2}\,\Be:\bm{\mathsf{C}}^\mathrm{dam}(\alpha):\Be& \quad \text{if \sf{open}},  \\
\frac{1}{2}(\bm{\varepsilon}-\bm{\varepsilon}^\mathrm{p}-\bm{\varepsilon}^\mathrm{r}):\bm{\mathsf{C}}:(\bm{\varepsilon}-\bm{\varepsilon}^\mathrm{p}-\bm{\varepsilon}^\mathrm{r}) + \frac{1}{2}\,\bm{\varepsilon}^\mathrm{p}:\bm{\mathsf{H}}^\mathrm{kin}(\alpha):\bm{\varepsilon}^\mathrm{p} & \quad \text{if \sf{closed}}. 
\end{dcases}
\label{c6:free}
\end{equation}
Equation \eqref{c6:free} shows that in the presence of open microcracks, frictional sliding is not expected, resulting in the exclusion of plastic and ratcheting strains from contributing to the free energy density. Consequently, the constitutive response in this regime depends solely on the total strain, and the model behaves in a brittle manner. Although previously accumulated plastic and ratcheting strains remain stored as internal variables, they do not influence the stress response while the material point remains in the open state. On the other hand, when microcracks are closed and sliding occurs, both the stored elastic energy and the blocked friction energy contribute to the strain energy density, and the plastic and ratcheting strains actively influence the constitutive response. The ratcheting strain is included in the elastic part of the free energy because it represents a slowly evolving, irreversible shift of the elastic reference state caused by microstructural rearrangements. In the rheological model (figure \ref{c6:ratch}), it acts as a slider in series with the elastic spring, altering the stored elastic energy at each cycle.


In equation \ref{c6:free}, the elasticity tensor $\bm{\mathsf{C}}$ is given by
\begin{equation}
\bm{\mathsf{C}}=K\bm{1}\otimes\bm{1}+2\mu\bigg(\bm{\mathsf{I}}-\frac{1}{3}\bm{1}\otimes\bm{1}\bigg),
\label{c3:Celas}
\end{equation}
where $K$ and $\mu$ are bulk and shear moduli. The damage-dependent elasticity tensor $\bm{\mathsf{C}}^\mathrm{dam}(\alpha)$ is decomposed to its volumetric and deviatoric parts as follows
\begin{equation}
\bm{\mathsf{C}}^\mathrm{dam}(\alpha)\coloneqq g_K(\alpha)K\bm{1}\otimes\bm{1}+2\,g_\mu(\alpha)\mu\bigg(\bm{\mathsf{I}}-\frac{1}{3}\bm{1}\otimes\bm{1}\bigg),
\label{c3:Cdam}
\end{equation}
in which $g_K(\alpha)$ and $g_\mu(\alpha)$ are the degradation functions defined as
\begin{equation}
g_K(\alpha)\coloneqq\frac{(1-\alpha)^2}{1+(b-1)[1-(1-\alpha)^2]}\, ,
\label{c3:gk}
\end{equation}
where $b$ is introduced as a dimensionless tuning parameter and $g_\mu(\alpha)$ is obtained as
\begin{equation}
g_\mu(\alpha)=\frac{g_K(\alpha)}{g_K(\alpha)+\frac{b_\mu}{b_K}[1-g_K(\alpha)]}\, .
\label{c3:gmu}
\end{equation}
The coefficients $b_K$ and $b_\mu$ are obtained using the Mori-Tanaka homogenization scheme \cite{mori1973a} as

\begin{equation}
b_K=\frac{16}{9}\frac{1-\nu^2}{1-2\nu}\,, \quad b_\mu=\frac{32}{45}\frac{(1-\nu)(5-\nu)}{2-\nu}\, .
\label{c3:ecopen}
\end{equation}
It is clear from equations \eqref{c3:gk} and \eqref{c3:gmu} that the degradation functions approach zero for fully damaged material ($\alpha=1$). The degradation functions $g_K$ and $g_\mu$ are introduced to describe the progressive reduction of the bulk and shear moduli due to microcrack growth, and are defined phenomenologically to reproduce the macroscopic stiffness degradation observed in quasi-brittle materials. However, the overall framework remains micromechanics-based, as the constitutive structure and the introduction of the internal strain variables originate from the homogenization of a representative volume element containing microcracks, following Eshelby-type solutions. Within this setting, the degradation functions act as macroscopic representations of the evolving microcracks.

The last function from the free energy formulation to be defined is the damage-dependent kinematic hardening modulus denoted as $\bm{\mathsf{H}}^\mathrm{kin}(\alpha)$, which is obtained meeting the continuity of free energy for the onset of microcrack closure when no ratcheting strain has yet accumulated, as
\begin{equation}
\bm{\mathsf{H}}^\mathrm{kin}(\alpha)=\Big[ {\bm{\mathsf{C}}^\mathrm{dam}}^{-1}(\alpha) - \bm{\mathsf{C}}^{-1} \Big]^{-1},
\label{c3:Hkin}
\end{equation}
which takes the decomposed form of
\begin{equation}
\bm{\mathsf{H}}^\mathrm{kin}(\alpha)= H^\mathrm{kin}_K(\alpha)\bm{1}\otimes\bm{1}+H^\mathrm{kin}_\mu(\alpha)\bigg(\bm{\mathsf{I}}-\frac{1}{3}\bm{1}\otimes\bm{1}\bigg),
\label{c3:decomhkin}
\end{equation}
where the volumetric and deviatoric parts are given by
\begin{equation}
H^\mathrm{kin}_K(\alpha)=\frac{g_K(\alpha)K}{1-g_K(\alpha)} \qquad \text{and} \qquad H^\mathrm{kin}_\mu(\alpha)=\frac{2\,g_\mu(\alpha)\mu}{1-g_\mu(\alpha)}\,.
\label{c3:HkinKmu}
\end{equation}
The kinematic hardening, as shown in equation \eqref{c3:Hkin}, is obtained from the stiffness tensor and is linked to the degradation function $g(\alpha)$. Consequently, there is no longer a necessity for phenomenological definitions of kinematic hardening.  The detailed derivation is presented in our previous work \cite{sarem2025a}. $\bm{\mathsf{H}}^\mathrm{kin}(\alpha)$, which here couples damage and plasticity, approaches infinity as the damage variable tends to zero. To prevent this, an initial damage is always required, representing the existence of microcracks which aligns with the physical assumptions of the micromechanics-based framework. The use of a small initial value has no effect on the final damage profile \cite{ulloa2022a}.

The stress-strain relations are obtained from the free energy density (equation \eqref{c6:free}) as
\begin{equation}
\bm{\sigma}(\bm{\varepsilon},\bm{\varepsilon}^\mathrm{p},\bm{\varepsilon}^\mathrm{r},\alpha)=\dfrac{\partial\psi}{\partial\bm{\varepsilon}}=\begin{dcases}
\bm{\mathsf{C}}^\mathrm{dam}(\alpha):\bm{\varepsilon}& \quad \text{if \sf{open}},  \\
\bm{\mathsf{C}}:(\bm{\varepsilon}-\bm{\varepsilon}^\mathrm{p}-\bm{\varepsilon}^\mathrm{r}) & \quad \text{if \sf{closed}}. 
\end{dcases}
\label{c6:sig}
\end{equation}
Furthermore, the generalized stresses conjugate to the plastic strain are obtained as
\begin{equation}
\Bsp(\bm{\varepsilon},\bm{\varepsilon}^\mathrm{p},\bm{\varepsilon}^\mathrm{r},\alpha)=-\frac{\partial\psi}{\partial\bm{\varepsilon}^\mathrm{p}}=\begin{dcases}
\bm{0}& \quad \text{if \sf{open}},  \\
\bm{\mathsf{C}}:(\bm{\varepsilon}-\bm{\varepsilon}^\mathrm{p}-\bm{\varepsilon}^\mathrm{r}) - \bm{\mathsf{H}}^\mathrm{kin}(\alpha):\bm{\varepsilon}^\mathrm{p} & \quad \text{if \sf{closed}}.
\end{dcases} 
\label{c6:sp}
\end{equation}
$\Bsp$ is interpreted as the local stress acting on microcrack faces within the RVE \cite{zhu2011a}. When microcracks are open, indicating no contact between them, $\Bsp$ tends to zero. Therefore, the following conditions are employed in the model to distinguish between open and closed microcracks or to differentiate the tensile and compressive/shear regimes at a macroscopic level as  
\begin{equation}
\begin{dcases}\tr\Bsp(\bm{\varepsilon},\bm{\varepsilon}^\mathrm{p},\alpha)=0 & \quad \text{if \sf{open}}, \\ \tr\Bsp(\bm{\varepsilon},\bm{\varepsilon}^\mathrm{p},\alpha) < 0 & \quad \text{if \sf{closed}}.\end{dcases}
\label{c3:trcon}
\end{equation}
Here, $\tr\Bsp$ is the volumetric component of the generalized stress $\Bsp$. Since the free energy also depends on the ratcheting strain, the generalized stresses conjugate to the ratcheting strains are obtained as follows
\begin{equation}
{s}^\mathrm{r}(\bm{\varepsilon},\bm{\varepsilon}^\mathrm{p},\bm{\varepsilon}^\mathrm{r},\alpha)=-\dfrac{\partial\psi}{\partial\bm{\varepsilon}^\mathrm{r}}=\begin{dcases}
\bm{0} & \quad \text{if \sf{open}},  \\
\bm{\mathsf{C}}:(\bm{\varepsilon}-\bm{\varepsilon}^\mathrm{p}-\bm{\varepsilon}^\mathrm{r}) & \quad \text{if \sf{closed}}. 
\end{dcases} 
\label{c6:sr}
\end{equation}
Finally, the generalized stresses conjugate to the crack phase-field are given by
\begin{equation}
{s}^\mathrm{d}(\bm{\varepsilon},\bm{\varepsilon}^\mathrm{p},\alpha)=-\dfrac{\partial\psi}{\partial\alpha}=\begin{dcases}
-\frac{1}{2}\,\bm{\varepsilon}:{\bm{\mathsf{C}}^\mathrm{dam}}'(\alpha):\bm{\varepsilon}& \quad \text{if \sf{open}},  \\
-\frac{1}{2}{\,\bm{\varepsilon}^\mathrm{p}:\bm{\mathsf{H}}^{\mathrm{kin}}}'(\alpha):\bm{\varepsilon}^\mathrm{p}  & \quad \text{if \sf{closed}}.  
\end{dcases} 
\label{c6:sd}
\end{equation}
It should be noted that the model employs a two-regime definition of the free energy, distinguishing between the open and closed microcrack states. This piecewise formulation intentionally reflects the unilateral mechanical behavior associated with crack opening and closure. In the original formulation, without ratcheting strain $\Ber$, the continuity of the free energy and its derivatives across the open–closed transition is enforced, leading to the definition of the kinematic hardening modulus in equation \eqref{c3:Hkin} \cite{ulloa2022a, sarem2023a}.\\
In the present formulation, the introduction of the ratcheting strain $\bm{\varepsilon}^\mathrm{r}$ results in a controlled, regime-switching discontinuity in the free energy. The switching condition between open and closed microcracks (see equation \eqref{c3:trcon}) ensures that this discontinuity occurs only at physically meaningful transitions, corresponding to the onset or loss of contact between crack faces. This discontinuity is therefore a representation of two distinct micromechanical states: open microcracks, where no contact or friction occurs, and closed microcracks, where sliding and ratcheting mechanisms are activated.\\
The following subsections show that dissipation is consistently formulated through convex, non-negative potentials, ensuring that the overall formulation remains thermodynamically admissible. Consequently, the Clausius–Duhem inequality is satisfied even in the presence of non-smooth energy branches. The accumulated ratcheting strain introduces an additional irreversible, dissipative mechanism without violating the variational structure of the model \cite{francfort2018a}.


\subsection{Plasticity criterion with a non-associated plastic flow rule}\label{ch3:plasevo}
To define the plastic evolution equations, the generalized stress $\Bsp$ is constrained to a non-empty, closed and convex set of admissible plastic generalized stresses $\mathbb{K}$
\begin{equation}
\Bsp\in\mathbb{K} \coloneqq \big\{\Bsp\in\mathbb{R}^{3\times 3}_{\mathrm{sym}} \ \ \vert \ \  f^\mathrm{p}(\Bsp)\leq 0\big\},
\label{c6:sc_setK}
\end{equation}
The choice of the plastic yield function $f^\mathrm{p}(\Bsp)$ depends on the selected plasticity model. The yield surface is a fundamental component of plasticity models that defines the boundary of the elastic domain \cite{ulloa2021b}. Here, the non-associative Drucker-Prager plasticity model is employed to capture frictional-dilational behavior of quasi-brittle materials for shear failure. Associative models predict excessive dilation for these materials which is unrealistic \cite{vermeer1998a}. From micromechanical point of view, plastic deformation in quasi-brittle materials is primarily due to the frictional sliding of microcracks. In highly porous materials, an additional contribution comes from pore collapse under compressive loading \cite{liu2022a}. The Drucker-Prager-type yield function to characterize frictional sliding in generalized stress space is defined as
\begin{equation}
f^\mathrm{p}(\bm{s}^\mathrm{p})\coloneqq\Vert\Bsp_{\mathrm{dev}}\Vert + \sqrt\frac{2}{3}A_\varphi \mathrm{tr} \Bsp,
\label{c3:fp}
\end{equation}
where $A_\varphi$ is a dimensionless friction coefficient. The set $\mathbb{K}$ represents a Drucker-Prager cone in generalized stress space. The apex of this cone is placed at the origin, meaning the model does not account for cohesion in generalized stress space. This ensures that $\Bsp_{\mathrm{dev}}$ can vanish during the opening-closing transition while still allowing for damage-dependent cohesion in true stress space \cite{ulloa2022a}.

Another key component of plasticity models is the evolution equation for plastic strain, governed by the flow rule, which determines how the plastic strain develops in response to stress once yielding occurs. In associative plasticity models, the flow rule is derived from the principle of maximum dissipation and corresponds to the normality rule, meaning the plastic strain rate is normal to the yield surface in stress space. In contrast, non-associative models abandon normality in favor of a more general flow rule, which is derived from a plastic potential different from the yield function \cite{ulloa2021b}. 

To incorporate ratcheting behavior into the flow rule, the plastic potential function $g^\mathrm{p}$ is defined as follows:
\begin{equation}
g^\mathrm{p}(\Bsp,\Bsr)\coloneqq\Vert\Bsp_{{\mathrm{dev}}}\Vert + \sqrt\frac{2}{3}(A_\theta \mathrm{tr}\Bsp+\beta_K A_\theta \mathrm{tr}\Bsr) + \beta_\mu \Vert\Bsr_{{\mathrm{dev}}}\Vert,
\label{c6:gp}
\end{equation}
where $A_\theta$ is the dilation coefficient which is assumed to take a value between zero and $A_\varphi$. $\beta_K$ and $\beta_\mu$ are ratcheting parameters associated with volumetric and deviatoric contributions, respectively. Since the underlying plasticity model is pressure-dependent, the ratcheting mechanism is also formulated to account for both volumetric and deviatoric strain components. In this context, macroscopic dilatancy is directly related to the normal opening of microcracks during frictional sliding, which arises from the roughness of the crack surfaces.

It is important to note that despite the non-associative nature of the plasticity model, it remains variational (see \cite{ulloa2021b} for further details). Moreover, the plastic potential is introduced at the macroscopic level and is not derived explicitly from micromechanical considerations. Rather, it provides a phenomenological representation of frictional dissipation associated with microcrack sliding, consistent with the overall thermodynamic framework.

The flow direction is defined by the subdifferential of the plastic potential:
\begin{equation}
\partial g(\bm{s}^\mathrm{p}, \bm{s}^\mathrm{r}) = \{ \hat{\bm{n}}, \hat{\bm{r}} \},
\end{equation}
with
\begin{equation}
\hat{\bm{n}}=\hat{\bm{n}}_\mathrm{dev} + \sqrt{\frac{2}{3}}A_\theta\bm{1}  \quad \text{and} \quad \hat{\bm{r}}=\beta_\mu\hat{\bm{r}}_\mathrm{dev} + \sqrt{\frac{2}{3}}\beta_K A_\theta\bm{1}, \quad \text{with} \squad \begin{cases}\hat{\bm{n}}_\mathrm{dev}\in\partial\Vert \bm{s}^\mathrm{p}_\mathrm{dev}\Vert, \\[4.5pt] \hat{\bm{r}}_{\,\mathrm{dev}}\in\partial\Vert \bm{s}^\mathrm{r}_\mathrm{dev}\Vert.\end{cases}
\label{c6:nhatrhat}
\end{equation}
Here, $\hat{\bm{n}}$ and $\hat{\bm{r}}$ represent the directions of plastic and ratcheting strain rates, respectively. Accordingly, the non-associative flow rule is defined as
\begin{equation}
\begin{aligned}
\{\dot{\bm{\varepsilon}}^\mathrm{p},\dot{\bm{\varepsilon}}^\mathrm{r}\}\in\mathbb{Q}(\bm{s}^\mathrm{p},\bm{s}^\mathrm{r})\coloneqq\big\{\lambda\,\{\hat{\bm{n}},\hat{\bm{r}}\}&\in\mathbb{R}^{3\times 3}_{\mathrm{dev}}\times\mathbb{R}^{3\times 3}_{\mathrm{dev}}  \\ &\vert \ \  \lambda\geq 0, \,\, \lambda= 0 \,\,\, \text{if} \,\,\, f^\mathrm{p}(\Bsp)<0\big\},
\end{aligned}
\label{c6:nonass}
\end{equation}
in which $\lambda$ is the plastic multiplier and $\mathbb{Q}(\Bsp)$ is the plastic flow set.

\subsection{Damage criterion with an associated flow rule}\label{ch3:damevo}
Similar to the work of Carrara et al. \cite{carrara2020a} and our previous work \cite{sarem2025a}, fatigue effects are considered by employing a fatigue variable $\mathcal{F}$ and a fatigue degradation function $h(\mathcal{F})$ with the following properties:
\begin{equation}
h(\mathcal{F}\leq\mathcal{F}_0)=1, \quad h(\mathcal{F}>\mathcal{F}_0)\in[0,1), \quad \frac{\mathrm{d}h(\mathcal{F})}{\mathrm{d}\mathcal{F}}\leq0,
\label{c5:fatdeg}
\end{equation}
in which $\mathcal{F}_0$ is a material threshold parameter, signifying that once the fatigue variable surpasses this threshold, fatigue-induced degradation initiates. This condition ensures that the fatigue degradation function decreases from one to zero and does not increase. The degradation function follows a phenomenological law and determines how fatigue impacts the fracture energy of the material \cite{li2023a}. Here, two categories of fatigue degradation functions are incorporated, namely asymptotic and logarithmic, to characterize the fatigue-induced damage. The asymptotic fatigue degradation function is given by
\begin{equation}
h(\mathcal{F})\coloneqq \begin{dcases} 1 & \text{if}  \quad \mathcal{F}(t) \leq \mathcal{F}_0, \\ 
\bigg(\frac{2\mathcal{F}_0}{\mathcal{F}(t)+\mathcal{F}_0}\bigg)^2 & \text{if} \quad \mathcal{F}(t) > \mathcal{F}_0,
\end{dcases}
\label{c5:asymp}
\end{equation}
and the logarithmic fatigue degradation function is given as follows
\begin{equation}
h(\mathcal{F})\coloneqq \begin{dcases} 1 & \text{if}  \quad \mathcal{F}(t) < \mathcal{F}_0, \\ 
\bigg[1-k\log\bigg(\frac{\mathcal{F}(t)}{\mathcal{F}_0}\bigg)\bigg]^2 & \text{if} \quad \mathcal{F}_0\leq\mathcal{F}(t) < \mathcal{F}_010^{1/k}, \\ 
0 & \text{if}  \quad \mathcal{F}(t) \geq \mathcal{F}_010^{1/k},
\end{dcases}
\label{c5:log}
\end{equation}
where $k$ is a material parameter that adjusts the slope of the logarithmic function. An increase in $k$ evidently amplifies the stress difference between the initial and final cycles. By choosing the latter degradation function, $k$ enhances the model's adaptability during calibration against experimental data, providing additional freedom to accurately represent the material's behavior \cite{sarem2023b}. Although alternative fatigue degradation functions have been proposed in the literature, it is important to highlight that these two functions have been validated \cite{carrara2020a} to effectively capture key aspects of the fatigue phenomenon, such as the relationship between load ratio (S) and number of cycles to failure (N), commonly represented by the $S-N$ curve.

In many studies, researchers have proposed degrading the fracture energy as the material accumulates strain \cite{alessi2018b} or energy \cite{carrara2020a, kristensen2020a, sarem2023b}, based on the understanding that cyclic loading, inelastic deformation, or microcrack accumulation progressively reduce the material’s ability to resist crack propagation. In this study, the fatigue history variable is defined to depend on the components of the free energy that are influenced by damage, following the approach proposed in \cite{carrara2020a}. The reasoning behind this choice is to ensure that only the energy contributing to damage evolution drives the degradation of fracture resistance. The fatigue variable $\mathcal{F}$ is thus defined as
\begin{equation}
\mathcal{F}(t) \coloneqq \int_{0}^t\dot{\vartheta}H\big(\dot{\vartheta}\big)\, \mathrm{d}\tau,
\label{c5:fatvar}
\end{equation}
in which $H$ represents the Heaviside function, preventing fatigue degradation during unloading and $\vartheta$ is the fatigue history variable considered as:
\begin{equation}
\vartheta(t)\coloneqq \begin{dcases} \frac{1}{2}\,\varepsilon:\bm{\mathsf{C}}^\mathrm{dam}(\alpha):\varepsilon \quad \text{if open}, \\
\frac{1}{2}\,\varepsilon^\mathrm{p}:\bm{\mathsf{H}}^{\mathrm{kin}}(\alpha):\varepsilon^\mathrm{p} \quad \text{if closed}. \end{dcases}
\label{c5:fathis}
\end{equation}
It is worth noting that the fatigue variable starts accumulating from the first cycle, which should be considered during calibration and parameter fitting \cite{kalina2023a}.

\begin{figure} [htb!]
\centering \includegraphics[scale=0.45]{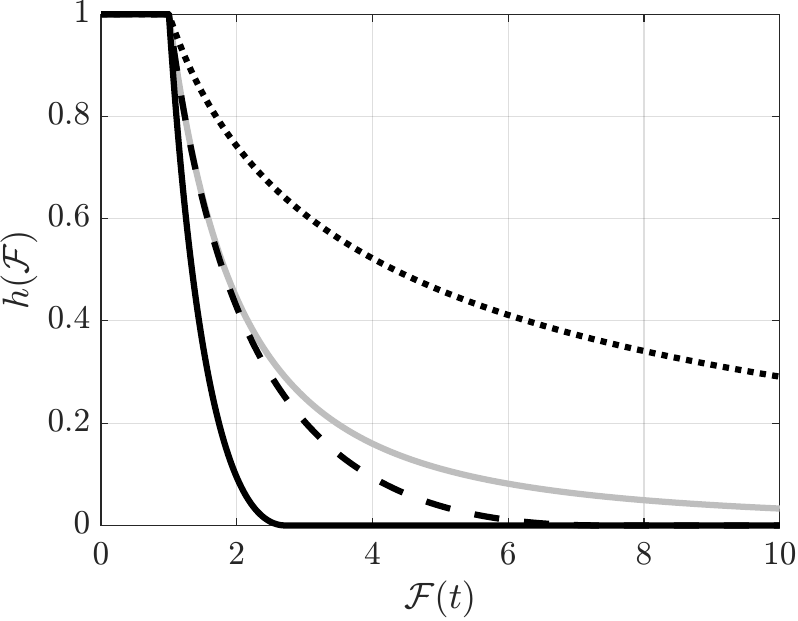}
\caption{Asymptotic degradation function (\textcolor{greylight}{$\full$}) and logarithmic degradation functions for $k=1$ ($\full$), $k=0.5$ ($\dashed$) and $k=0.2$ ($\dotted$).}
\label{fig5:fatdeg}
\end{figure}

Figure \ref{fig5:fatdeg} compares the two degradation functions and illustrates their evolution with respect to the fatigue history variable, assuming a fatigue threshold of $\mathcal{F}_0=1$. The influence of different values of the parameter $k$ is shown for the logarithmic function. The asymptotic function shows a smooth and gradual decrease, while the logarithmic function displays a sharper degradation, with the rate of decay increasing for larger values of $k$. The asymptotic function approaches zero asymptotically but never reaches total degradation, whereas the logarithmic function reduces to zero at a finite fatigue level. In the logarithmic case, the parameter $k$ controls the sharpness of the drop, while the asymptotic function lacks such a parameter, resulting in a fixed decay rate.

The generalized stress constraint is defined in a non-empty, closed and convex set of admissible damage generalized stresses $\mathbb{K}^\mathrm{d}$
\begin{equation}
s^\mathrm{d}\in\mathbb{K}^\mathrm{d} \coloneqq \big\{s^\mathrm{d}\in\mathbb{R}_{+} \ \ \vert \ \  f^\mathrm{d}(s^\mathrm{d};\Bsp)\leq 0\big\}.
\label{c3:sc_setKd}
\end{equation}
The damage yield function $f^\mathrm{d}$ is defined within the framework of irreversible thermodynamics as
\begin{equation}
f^\mathrm{d}(s^\mathrm{d};\Bsp)\coloneqq s^\mathrm{d}-h(\mathcal{F})\frac{{G}_\mathrm{c}(\Bsp)\alpha}{{\ell}}+{\ell}\,\mathrm{div}[h(\mathcal{F}){G}_\mathrm{c}(\Bsp)\nabla\alpha].
\label{c5:fd}
\end{equation}
where ${\ell}$ is the length scale of the phase-field model which regularizes the displacement discontinuities due to cracks. Here, the degradation function acts as a modulation parameter for dissipation without changing the fundamental structure of the micromechanics-based phase-field model. Furthermore, the fracture energy ${G}_\mathrm{c}$ is chosen as
\begin{equation}
{G}_\mathrm{c}(\bm{s}^\mathrm{p})\coloneqq\begin{dcases}
G_\mathrm{cI}  \quad &\text{if }  \tr\bm{s}^\mathrm{p}=0, \\
G_\mathrm{cII} \quad &\text{if }  \tr\bm{s}^\mathrm{p}<0.
\end{dcases}
\label{c3:Gcsp}
\end{equation}
This distinction allows mode I fracture energy $G_\mathrm{cI}$ to govern tensile fractures, while mode II fracture energy $G_\mathrm{cII}$ accounts for compressive and shear fractures. Although the discontinuity in the fracture energy definition does not violate the variational formulation or thermodynamics consistency, it brings numerical issues due to the abrupt change in its value.

\begin{figure} [htb!]
\centering
\includegraphics[scale=0.6]{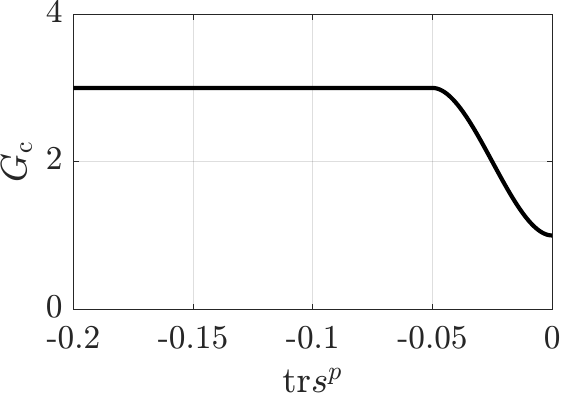}
\caption{Regularized fracture energy.}
\label{fig3:reg}
\end{figure}

To address this challenge, a mathematical regularization is introduced to smooth the transition between tensile and shear fracture modes. The regularized function that replaces equation \eqref{c3:Gcsp} is defined as
\begin{equation}
G_\mathrm{c}(\bm{s}^\mathrm{p}) \coloneqq
\begin{cases}
G_\mathrm{cI} &\text{if }  \tr \bm{s}^\mathrm{p} = 0, \\[6pt]
G_\mathrm{cII} + (G_\mathrm{cI} - G_\mathrm{cII}) \cdot s(\xi) &\text{if }  -\epsilon < \tr \bm{s}^\mathrm{p} < 0, \\[6pt]
G_\mathrm{cII} &\text{if }  \tr \bm{s}^\mathrm{p} \leq -\epsilon.
\end{cases}
\label{c3:Gc_smooth}
\end{equation}
Here, $\epsilon > 0$ controls the width of the transition zone. The function $s(\xi)$ is known as the smoothstep function, a cubic polynomial interpolant that provides a smooth transition. The normalized parameter $\xi$ reparametrizes $\tr \Bsp$ within the smoothing interval and is defined as follows \cite{stoer1980a}:
\begin{equation}
\xi = \frac{\tr \bm{s}^\mathrm{p} + \epsilon}{\epsilon}, \quad
s(\xi) = 3\xi^2 - 2\xi^3.
\label{c3:smoothstep}
\end{equation}
This regularization ensures that the fracture energy function is continuous and at least $C^1$-differentiable across the transition zone, improving numerical robustness. The behavior of equation \eqref{c3:Gc_smooth} is illustrated in figure \ref{fig3:reg}, assuming $G_\mathrm{cI}=1$, $G_\mathrm{cII}=3$ and $\epsilon=0.05$.
%

\subsection{Dissipation potential}\label{ch3:dispot}
The dissipation potential characterizes the energy dissipated during irreversible processes such as damage, plastic deformation and ratcheting. It is defined as a function of the rates of internal variables. A thermodynamically admissible formulation of the dissipation potential is expressed as the sum of plastic and damage contributions \cite{alessi2019a}:
\begin{equation}
{\phi(\dot{\bm{\varepsilon}}^\mathrm{p},\dot{\bm{\varepsilon}}^\mathrm{r},\dot{\alpha},\nabla\dot{\alpha};\alpha,\nabla\alpha,{\bm{s}^\mathrm{p}},{\bm{s}^\mathrm{r}})=\phi^\mathrm{p}(\dot{\bm{\varepsilon}}^\mathrm{p},\dot{\bm{\varepsilon}}^\mathrm{r};{\bm{s}^\mathrm{p}},{\bm{s}^\mathrm{r}})+\phi^\mathrm{d}(\dot{\alpha},\nabla\dot{\alpha};\alpha,\nabla\alpha,{\bm{s}^\mathrm{p}}),}
\label{c3:phi}
\end{equation}
where $\phi^\mathrm{p}$ and $\phi^\mathrm{d}$ represent plastic and damage dissipation potentials, respectively. 

To define the plastic dissipation and incorporate ratcheting effects, the set of admissible stress states is first specified. The non-associative model maintains a variational structure if a state-dependent convex set is defined as follows \cite{francfort2018a}:
\begin{equation}
\mathbb{L}(\Bsp,\Bsr)\coloneqq\bigg\{\{\tilde{\bm{s}}^\mathrm{p},\tilde{\bm{s}}^\mathrm{r}\}\in\mathbb{R}^{3\times 3}_{\mathrm{sym}}\times \mathbb{R}^{3\times 3}_{\mathrm{sym}} \ \ \vert \ \ 
g^\mathrm{p}(\tilde{\bm{s}}^\mathrm{p},\tilde{\bm{s}}^\mathrm{r})\leq g^\mathrm{p}(\Bsp,\Bsr)-f^\mathrm{p}(\Bsp) \bigg\},
\label{c6:lsp}
\end{equation}
where $\tilde{\bm{s}}^\mathrm{p}$ and $\tilde{\bm{s}}^\mathrm{r}$ denote admissible variations of $\Bsp$ and $\Bsr$, respectively. Based on this set, and following the principle of maximum dissipation \cite{hill1948a, moreau1974a}, the plastic dissipation potential is defined as:
\begin{equation}
\phi^\mathrm{p}(\dot{\bm{\varepsilon}}^\mathrm{p},\dot{\bm{\varepsilon}}^\mathrm{r};\Bsp,\Bsr)=\mathrm{sup}\big\{\tilde{\bm{s}}^\mathrm{p}:\dot{\bm{\varepsilon}}^\mathrm{p} +  {\tilde{\bm{s}}^\mathrm{r}}:\dot{\bm{\varepsilon}}^\mathrm{r} - I_{\mathbb{L}(\bm{s}^\mathrm{p},\bm{s}^\mathrm{r})}(\tilde{\bm{s}}^\mathrm{p},\tilde{\bm{s}}^\mathrm{r})\big\}.
\label{c6:disspotnona}
\end{equation}
This means that for a given plastic strain rate $\dot{\bm{\varepsilon}}^\mathrm{p}$ and ratcheting strain rate $\dot{\bm{\varepsilon}}^\mathrm{r}$, the plastic dissipation potential is the supremum (upper bound) of the work done by any admissible plastic stress $\tilde{\bm{s}}^\mathrm{p}$ and ratcheting stress $\tilde{\bm{s}}^\mathrm{r}$.
From equations \eqref{c6:lsp} and \eqref{c6:disspotnona}, the state-dependent plastic dissipation potential is evaluated for all $\{\dot{\bm{\varepsilon}}^\mathrm{p},\dot{\bm{\varepsilon}}^\mathrm{r}\} \in\mathbb{R}^{3\times 3}_{\mathrm{sym}}\times \mathbb{R}^{3\times 3}_{\mathrm{sym}}$
\begin{equation}
\begin{aligned}
\phi^\mathrm{p}(\dot{\bm{\varepsilon}}^\mathrm{p},\dot{\bm{\varepsilon}}^\mathrm{r};\Bsp,\Bsr)=\sup\bigg\{{\tilde{\bm{s}}^\mathrm{p}}:\dot{\bm{\varepsilon}}^\mathrm{p} + {\tilde{\bm{s}}^\mathrm{r}}:\dot{\bm{\varepsilon}}^\mathrm{r} \ \ &\vert \  \ \Vert\tilde{\bm{s}}^\mathrm{p}_{{\mathrm{dev}}}\Vert + \sqrt\frac{2}{3}(A_\theta \mathrm{tr}\tilde{\bm{s}}^\mathrm{p}+\beta_K A_\theta \mathrm{tr}\tilde{\bm{s}}^\mathrm{r})+ \beta_\mu\Vert\tilde{\bm{s}}^\mathrm{r}_{{\mathrm{dev}}}\Vert \\ 
&\hspace*{-0cm} \leq \sqrt{\frac{2}{3}}(A_\theta-A_\varphi)\tr\bm{s}^\mathrm{p} +\sqrt{\frac{2}{3}}A_\theta \beta_K \tr\Bsr +\beta_\mu \Vert\Bsr_{{\mathrm{dev}}}\Vert \bigg\}.
\end{aligned}
\label{c6:disspot1}
\end{equation}
By decomposing all stress and strain terms into their deviatoric and volumetric parts, this expression becomes:
\begingroup
\allowdisplaybreaks
\begin{equation}
\begin{aligned}
\phi^\mathrm{p}(\dot{\bm{\varepsilon}}^\mathrm{p},\dot{\bm{\varepsilon}}^\mathrm{r};\Bsp,\bm{s}^\mathrm{r})
&=\sup\bigg\{{\tilde{\bm{s}}_{\mathrm{dev}}^\mathrm{p}}:\dot{\bm{\varepsilon}}_{\mathrm{dev}}^\mathrm{p}+{\tilde{\bm{s}}_{\mathrm{dev}}^\mathrm{r}}:\dot{\bm{\varepsilon}}_{\mathrm{dev}}^\mathrm{r}+\tfrac{1}{3}(\tr{\tilde{\bm{s}}^\mathrm{p}}\tr \dot{\bm{\varepsilon}}^\mathrm{p} + \tr{\tilde{\bm{s}}^\mathrm{r}}\tr\dot{\bm{\varepsilon}}^\mathrm{r})\ \ \vert  \\
&\hspace*{1.5cm} \tr\tilde{\bm{s}}^\mathrm{p}\leq \frac{(A_\theta-A_\varphi)}{A_\theta}\tr\bm{s}^\mathrm{p}-\sqrt{\frac{3}{2}}\frac{1}{A_\theta}\Vert\tilde{\bm{s}}^\mathrm{p}_{{\mathrm{dev}}}\Vert + \beta_K\tr\Bsr \\
&\hspace*{2.5cm} + \sqrt{\frac{3}{2}}\frac{\beta_\mu}{A_\theta} \Vert \Bsr_{\mathrm{dev}} \Vert -\sqrt{\frac{3}{2}}\frac{\beta_\mu}{A_\theta}\Vert\tilde{\bm{s}}^\mathrm{r}_{{\mathrm{dev}}}\Vert - \beta_K\tr\tilde{\bm{s}}^\mathrm{r}\bigg\}
\\
&=\sup\bigg\{\dfrac{\tr\dot{\bm{\varepsilon}}^\mathrm{p}}{3A_\theta}\bigg((A_\theta-A_\varphi)\tr\bm{s}^\mathrm{p}+\sqrt{\frac{2}{2}}\beta_\mu \Vert\Bsr_{\mathrm{dev}}\Vert+\beta_K A_\theta \mathrm{tr}\Bsr\bigg) \\
&\hspace*{1.2cm}\Vert{\tilde{\bm{s}}_{\mathrm{dev}}^\mathrm{p}}\Vert\bigg(\Vert\dot{\bm{\varepsilon}}_{\mathrm{dev}}^\mathrm{p}\Vert - \frac{1}{\sqrt{6}A_\theta}\tr\dot{\bm{\varepsilon}}^\mathrm{p}\bigg)+\frac{1}{3}\tr\tilde{\bm{s}}^\mathrm{r}(\tr\dot{\bm{\varepsilon}}^\mathrm{r}-\beta_K\tr\dot{\bm{\varepsilon}}^\mathrm{p})\\
&\hspace*{5.8cm}+\Vert\tilde{\bm{s}}_{\mathrm{dev}}^\mathrm{r}\Vert\bigg(\Vert \dot{\bm{\varepsilon}}^\mathrm{r}_\mathrm{dev} \Vert-\frac{\beta_\mu}{\sqrt{6}A_\theta}\tr\dot{\varepsilon}^\mathrm{p}\bigg)
\bigg\}.
\end{aligned}
\label{c6:disspot2}
\end{equation}
\endgroup
It is important to note that $\Vert{\tilde{\bm{s}}_{\mathrm{dev}}^\mathrm{p}}\Vert$ and $\Vert{\tilde{\bm{s}}_{\mathrm{dev}}^\mathrm{r}}\Vert$ are free variables, while the terms multiplying them i.e., $(\Vert\dot{\bm{\varepsilon}}_{\mathrm{dev}}^\mathrm{p}\Vert - \frac{1}{\sqrt{6}A_\theta}\tr\dot{\bm{\varepsilon}}^\mathrm{p})$ and $(\Vert \dot{\bm{\varepsilon}}^\mathrm{r}_\mathrm{dev} \Vert-\frac{\beta_\mu}{\sqrt{6}A_\theta}\tr\dot{\varepsilon}^\mathrm{p})$, must be bounded to ensure a finite supremum.
These conditions lead directly to the final form of the plastic dissipation potential:
\begin{equation}
\begin{aligned}
\phi^\mathrm{p}(&\dot{\bm{\varepsilon}}^\mathrm{p},\dot{\bm{\varepsilon}}^\mathrm{r};\bm{s}^\mathrm{p},\bm{s}^\mathrm{r})=\\&
\hspace*{1cm}\begin{dcases}
\frac{\tr\dot{\bm{\varepsilon}}^\mathrm{p}}{3A_\theta}\bigg((A_\theta-A_\varphi)\tr\bm{s}^\mathrm{p} \\[-1cm] \hspace{1cm} +\sqrt{\frac{2}{2}}\beta_\mu \Vert\Bsr_{\mathrm{dev}}\Vert+\beta_K A_\theta \mathrm{tr}\Bsr\bigg) \quad  \text{if} \ \begin{dcases} \tr\dot{\bm{\varepsilon}}^\mathrm{p}\geq\sqrt{6}A_\theta\Vert \dot{\bm{\varepsilon}}^\mathrm{p}_\mathrm{dev}\Vert, \\ \Vert \dot{\bm{\varepsilon}}^\mathrm{r}_\mathrm{dev} \Vert \leq   \frac{\beta_\mu}{\sqrt{6}A_\theta}\tr\dot{\varepsilon}^\mathrm{p}, \\ \tr\dot{\bm{\varepsilon}}^\mathrm{r}=\beta_K\tr\dot{\bm{\varepsilon}}^\mathrm{p},\end{dcases} \\
+\infty \quad \text{otherwise}.
\end{dcases} 
\end{aligned}
\label{c6:disspot3}
\end{equation}
The derived extended plastic dissipation potential effectively captures the progressive and irreversible nature of ratcheting under cyclic loading. The explicit control provided by the ratcheting parameters $\beta_\mu$ and $\beta_K$ enables the simulation of complex cyclic behavior, such as mean strain accumulation and asymmetric hysteresis loops.

The damage dissipation potential is derived following the same procedure as the plastic dissipation potential, starting from the generalized stress constraint (equation \eqref{c3:sc_setKd}), which is governed by the damage yield condition (equation \eqref{c5:fd}). Once again, by applying the principle of maximum dissipation to damage evolution and considering the damage flow rule:
\begin{equation}
\dot{\alpha}\in \partial I_{\mathbb{K}^\mathrm{d}}(s^\mathrm{d}),
\label{c3:flow_d}
\end{equation} 
the damage dissipation potential is defined in a variational form as:
\begin{equation}
\begin{aligned}
\int_\Omega\phi^\mathrm{d}(\dot{\alpha},\nabla\dot{\alpha};\alpha,\nabla\alpha,{\Bsp})\,\mathrm{d}\bm{x}
&=\mathrm{sup}\bigg\{\int_\Omega\Big(\tilde{s}^\mathrm{d}\,\dot{\alpha} - I_{\mathbb{K}^\mathrm{d}}(\tilde{s}^\mathrm{d})\Big)\,\mathrm{d}\bm{x}\bigg\} \\
&= \sup \bigg\{ \int_{\Omega}  \Big(\tilde{s}^\mathrm{d} \dot{\alpha} \ \vert \ \tilde{s}^\mathrm{d} \leq h(\mathcal{F})\frac{G_\mathrm{c} (\bm{s}^\mathrm{p})}{\ell} \alpha - \ell \operatorname{div} \big[h(\mathcal{F}) G_\mathrm{c} (\bm{s}^\mathrm{p}) \nabla \alpha \big] \Big) \, \mathrm{d}\bm{x} \bigg\}.
\end{aligned}
\label{c3:appvarphid}
\end{equation}
By substituting the upper bound of $s^{\mathrm{d}}$, the following is obtained:
\begin{equation}
\int_\Omega\phi^\mathrm{d}\,\mathrm{d}\bm{x}
= \sup \bigg\{ \int_{\Omega} \Big(h(\mathcal{F})\frac{G_\mathrm{c} (\bm{s}^\mathrm{p})}{\ell} \alpha \dot{\alpha} - \ell \operatorname{div} \big[h(\mathcal{F}) G_\mathrm{c} (\bm{s}^\mathrm{p}) \nabla \alpha \big] \dot{\alpha} \Big) \, \mathrm{d}\bm{x} \bigg\}.
\label{c3:appvarphid1}   
\end{equation}
Applying the product rule to the second term in the integral:
\begin{equation}
\int_{\Omega} \operatorname{div} \big[h(\mathcal{F}) G_\mathrm{c} (\bm{s}^\mathrm{p}) \nabla \alpha \big] \dot{\alpha} \, \mathrm{d}\bm{x} = 
\int_{\Omega} \operatorname{div} \big[ \dot{\alpha} h(\mathcal{F}) G_\mathrm{c} (\bm{s}^\mathrm{p}) \nabla \alpha \big] \, \mathrm{d}\bm{x} - \int_{\Omega} h(\mathcal{F}) G_\mathrm{c} (\bm{s}^\mathrm{p}) \nabla \alpha \cdot \nabla \dot{\alpha} \, \mathrm{d}\bm{x}.
\label{c3:prorul} 
\end{equation}
Next, by employing the divergence theorem:
\begin{equation}
\int_{\Omega} \operatorname{div} \big[\dot{\alpha} h(\mathcal{F}) G_\mathrm{c} (\bm{s}^\mathrm{p}) \nabla \alpha\big] \, \mathrm{d}\bm{x} = 
\int_{\partial \Omega} (h(\mathcal{F}) G_\mathrm{c} (\bm{s}^\mathrm{p}) \nabla \alpha) \cdot \mathbf{n} \, \mathrm{d}S.
\label{c3:divtheo} 
\end{equation}
Applying the natural boundary condition, which enforces zero flux of $\alpha$ at the boundary:
\begin{equation}
(h(\mathcal{F}) G_\mathrm{c} (\bm{s}^\mathrm{p}) \nabla \alpha) \cdot \mathbf{n} = 0 \quad \text{on } \partial \Omega,
\label{c3:natboun} 
\end{equation}
the equation \eqref{c3:prorul} simplifies to
\begin{equation}
\int_{\Omega} \operatorname{div} \big[h(\mathcal{F}) G_\mathrm{c} (\bm{s}^\mathrm{p}) \nabla \alpha \big] \dot{\alpha} \, \mathrm{d}\bm{x} = - \int_{\Omega} h(\mathcal{F}) G_\mathrm{c} (\bm{s}^\mathrm{p}) \nabla \alpha \cdot \nabla \dot{\alpha} \, \mathrm{d}\bm{x}.
\label{c3:prorul2} 
\end{equation}
Substituting this result back into equation \eqref{c3:appvarphid1}, the following expression is obtained:
\begin{equation}
\int_\Omega\phi^\mathrm{d}\,\mathrm{d}\bm{x}
= \sup \bigg\{ \int_{\Omega} \Big(h(\mathcal{F})\frac{G_\mathrm{c} (\bm{s}^\mathrm{p})}{\ell} \alpha \dot{\alpha} - \ell h(\mathcal{F}) G_\mathrm{c} (\bm{s}^\mathrm{p}) \nabla \alpha \nabla \dot{\alpha} \Big) \, \mathrm{d}\bm{x} \bigg\}.
\label{c3:appvarphid2}   
\end{equation}
Finally, the explicit form of the damage dissipation potential is obtained as follows:
\begin{equation}
\phi^\mathrm{d}(\dot{\alpha},\nabla\dot{\alpha};\alpha,\nabla\alpha,{\bm{s}^\mathrm{p}})\coloneqq\begin{dcases}h(\mathcal{F})\frac{{G}_\mathrm{c}(\bm{s}^\mathrm{p})}{{\ell}}\big(\alpha\,\dot{\alpha}+{\ell}^{2}\nabla\alpha\cdot\nabla\dot{\alpha}\big) \quad \text{if } \dot{\alpha}\geq 0 ,\\
+\infty \quad \text{otherwise}.\end{dcases} 
\label{c3:phi_d}
\end{equation}
The damage dissipation potential includes a fatigue degradation function $h(\mathcal{F})$, which reflect the progressive weakening of the material under cyclic loading. This reduces the effective fracture energy and modifies both the local and gradient contributions of damage evolution. The condition $\dot{\alpha}\geq0$ enforces the irreversibility of damage evolution.

\section{Numerical implementation}\label{ch6:numimp}
This section presents the numerical implementation of the micromechanics-based phase-field model. A three-field staggered algorithm is adopted to decouple the coupled system of PDEs in order to solve a sequence of convex sub-problems. These sub-problems are obtained by minimizing the energy functional with respect to one field variable at a time \cite{miehe2010b, wambacq2021b}. Through the staggered solution technique, equilibrium, damage and plasticity equations are solved independently for the displacement field $\Bu$, the crack phase-field $\alpha$, and the plastic and ratcheting strains $\{\Bep,\Ber\}$, using Newton-Raphson schemes.  While the energy functional remains non-convex as a whole, it is convex with respect to each individual variable \cite{alessi2013a}.

Algorithm \ref{algorithm3:stag} provides an overview of the staggered procedure employed in this study. To begin with, the nonlinear mechanical equilibrium equation is solved to find $\Bu$. Subsequently, the damage evolution problem is solved to obtain $\alpha$ using the \emph{history-field} approach \cite{miehe2010b}. Next, a return-mapping algorithm is employed to update $\Bep$ and $\Ber$. In this sub-problem, the algorithm determines the state of the solid matrix, denoted as \emph{state}, in terms of open or closed microcracks using equation \eqref{c3:trcon}.  Through this iterative procedure, the governing equations are solved at each loading step and for every Gauss point, ensuring a consistent update of the field variables. The detailed solution procedure for the three sub-problems is provided in the next subsections.

The state variables at previous time steps, from $0$ to $t_n$ are assumed known and are used to determine the variables at the next time step $t_{n+1}$. For clarity, the following notations are used throughout this section: a quantity $\Box$ evaluated at time step $t_n$ is denoted as $\Box_n$, while a quantity at $t_{n+1}$ is written without a subscript, i.e., $\Box\coloneqq \Box_{n+1}$. Additionally, the increment of a quantity from $t_n$ to $t_{n+1}$ is defined as $\Delta\Box \coloneqq \Box - \Box_n$. These notations provide a consistent framework for describing the numerical implementation.

\begin{algorithm}
\hspace*{\algorithmicindent} \textbf{Input}: primary fields at the previous time step $\Bvu_n$, $\Bvep_n$, $\Ber_n$, $\alpha_n$ and $\mathrm{{state}_n}$. \\
\hspace*{\algorithmicindent} \textbf{Output}: primary fields at the current time step $\Bvu$, $\Bvep$, $\Ber$, $\alpha$ and $\mathrm{state}$. 
\begin{algorithmic}[1]
\State Initialize iterations with $j\coloneqq 0$ and set $\{\Bvu^{(0)},\bm{\varepsilon}^{\mathrm{p}(0)},\bm{\varepsilon}^{\mathrm{r}(0)},\alpha^{(0)},\mathrm{{state}^{(0)}}\}\coloneqq \{\Bvu_n,\bm{\varepsilon}^{\mathrm{p}}_n,\bm{\varepsilon}^{\mathrm{r}}_n,\alpha_n,\mathrm{{state}_n}\}$.
\Repeat  
\State Set $j \leftarrow j+1$.
\State Solve the \underline{mechanical balance equation} for $\bm{u}^{(j)}$ using $\{\bm{\varepsilon}^{\mathrm{p}(j-1)},\bm{\varepsilon}^{\mathrm{r}(j-1)},\alpha^{(j-1)},\mathrm{{state}^{(j-1)}}\}$.
\State Solve the non-linear \underline{damage evolution problem} for $\alpha^{(j)}$ using $\{\bm{u}^{(j)},\bm{\varepsilon}^{\mathrm{p}(j-1)},\bm{\varepsilon}^{\mathrm{r}(j-1)},\mathrm{{state}^{(j-1)}}\}$.
\State Solve the non-linear \underline{plastic and ratcheting evolution problem} for $\{\bm{\varepsilon}^{\mathrm{p}(j)},\bm{\varepsilon}^{\mathrm{r}(j)},\mathrm{{state}^{(j)}}\}$ using $\{\bm{u}^{(j)},\alpha^{(j)}\}$.
\State Update $$r_{\Bvu}^{
(j)}\coloneqq \int_\Omega  \Big[\Bvsig\big(\nabla^{\mathrm{s}}\Bvu^{(j)},\bm{\varepsilon}^{\mathrm{p}(j)},\bm{\varepsilon}^{\mathrm{r}(j)},\alpha^{(j)}\big):\nabla^\mathrm{s}\tilde{\Bvu} - \rho\bm{b}\cdot\tilde{\bm{u}}\Big]\,\mathrm{d}\bm{x} - \int_{\partial\Omega_\mathrm{N}}\bar{\bm{t}}\cdot\tilde{\bm{u}}\,\mathrm{d}S \quad \forall\,\tilde{\Bvu}\in\tilde{\mathscr{U}}.$$ and
$$r_{\alpha}^{(j)}  \eqqcolon  \int_\Omega \bigg[ s^\mathrm{d\,hist}(\bm{\varepsilon}^{(j)},{\bm{\varepsilon}^\mathrm{p}}^{(j)},\alpha^{(j)})\,\tilde{\alpha}  - h(\mathcal{F})\frac{{G}_\mathrm{c}(\bm{s}^\mathrm{p})}{{\ell}}\big(\alpha^{(j)}\,\tilde{\alpha}+{\ell}^{2}\nabla\alpha^{(j)}\cdot\nabla\tilde{\alpha}\big) \bigg] \,\mathrm{d}\bm{x} \quad \forall\,\tilde{\alpha}\in\mathrm{H}^1(\Omega;\mathbb{R}).$$  
\Until $\big\vert r_{\Bvu}^{(j)}\big\vert\leq \mathtt{TOL}_{\Bvu}$ and $\big\vert r_{\alpha}^{(j)}\big\vert\leq \mathtt{TOL}_{\alpha}$.
\State Set $\{\Bvu,\Bvep,\Ber,\alpha,\mathrm{state}\}\coloneqq\{\Bvu^{(j)},\bm{\varepsilon}^{\mathrm{p}(j)},\bm{\varepsilon}^{\mathrm{r}(j)},\alpha^{(j)},\mathrm{{state}^{(j)}}\}$.
\end{algorithmic}
\caption{The staggered solution procedure.}\label{algorithm3:stag}
\end{algorithm} 

\subsection{Mechanical balance equation}\label{ch3:mechbaleq}
First, the equilibrium problem is solved for the displacement $\Bu$ by linearizing the mechanical balance equation, while the other three variables $\Bep$, $\Ber$ and $\alpha$ are fixed. The weak form of the mechanical balance is given by
\begin{equation}
\int_\Omega \big( \Bsig:\nabla^\mathrm{s}\tilde{\Bu}-\rho\bm{b}\cdot\tilde{\bm{u}}\big)\,\mathrm{d}\bm{x} - \int_{\partial\Omega_\mathrm{N}}\bar{\bm{t}}\cdot\tilde{\bm{u}}\,\mathrm{d}S =0 \quad \forall\,\tilde{\Bu}\in\tilde{\mathscr{U}}.
\label{c3:weaku}
\end{equation}
The given equation involves the body forces per unit mass denoted by $\bm{b}$, the unit mass denoted by $\rho$, and the imposed tractions denoted by $\bar{\bm{t}}$. When linearized, the equation leads to the following expression
\begin{equation}
\begin{aligned}
&\int_\Omega \nabla^\mathrm{s}[\Bu^{(k+1)}-\Bu^{(k)}]:\bm{\mathsf{C}}^{\mathrm{ep}(k)}:\nabla^\mathrm{s}\tilde{\Bu}\,\mathrm{d}\bm{x} \\ =&-\int_\Omega  \big(\Bsig^{(k)}:\nabla^\mathrm{s}\tilde{\Bu}  - \rho\bm{b}\cdot\tilde{\bm{u}}\big)\,\mathrm{d}\bm{x} + \int_{\partial\Omega_\mathrm{N}}\bar{\bm{t}}\cdot\tilde{\bm{u}}\,\mathrm{d}S \eqqcolon -r_{\Bu}^{(k)} \quad \forall\,\tilde{\Bu}\in\tilde{\mathscr{U}}.
\end{aligned}
\label{c3:weakualg}
\end{equation}
It is noted that equation \ref{c3:weakualg} is linear with respect to the displacement increment within a given iteration. However, since the stress and the consistent tangent operator depend on the evolving internal variables, the overall equilibrium problem is nonlinear and is therefore solved using a Newton-type iterative scheme. Specifically, the equation is solved iteratively, where each iteration produces a displacement denoted by $\Bu^{(k+1)}$. The process continues until the residual, represented by $r_{\Bu}^{(k)}$, is smaller than a predefined tolerance $\mathtt{TOL}_{\Bu}$.  In this context, the index $k$ indicates the current iteration counter. The terms $\bm{\mathsf{C}}^{\mathrm{ep}(k)}\coloneqq\partial\Bsig^{(k)}/ \partial \Be^{(k)}$ and $\Bsig^{(k)}$ refer to the consistent tangent and stress tensor at the current iteration \cite{deborst2012a}, respectively.

\subsection{Damage evolution problem}\label{ch3:damevo}
The equilibrium is solved in the previous step to obtain the displacement $\Bu$. The subsequent step in the staggered solution approach is to address the following variational inequality to obtain $\alpha$
\begin{equation}
\begin{aligned}
\int_\Omega \bigg( -s^\mathrm{d}(\bm{\varepsilon},\bm{\varepsilon}^\mathrm{p},\alpha)\,\tilde{\alpha}  +h(\mathcal{F}) \frac{{G}_\mathrm{c}(\bm{s}^\mathrm{p})}{{\ell}}\big(\alpha\,\tilde{\alpha}+{\ell}^{2}\nabla\alpha\cdot\nabla\tilde{\alpha}\big) + \partial I_{\mathbb{R}_+}(\Delta \alpha)\,\tilde{\alpha} \bigg) \,\mathrm{d}\bm{x} \ni0 \quad \forall\,\tilde{\alpha}\in\mathrm{H}^1(\Omega;\mathbb{R}).
\end{aligned}
\label{c3:weakd}
\end{equation}
The incremental expression of the irreversibility condition is given by
\begin{equation}
\Delta\alpha\geq 0.
\label{c3:irr}
\end{equation}
The indicator function is employed in the aforementioned equation to ensure that the damage variable remains irreversible. The subdifferential of the indicator function is defined as
\begin{equation}
\partial I_{\mathbb{R}_+}(\Delta \alpha) =
\begin{cases}
0, & \text{if } \Delta \alpha > 0, \\
\mathbb{R}_-, & \text{if } \Delta \alpha = 0, \\
\varnothing, & \text{if } \Delta \alpha < 0.
\end{cases}
\label{c3:subind}
\end{equation}
The damage driving force is calculated assuming that the opening/closure state (equation \eqref{c3:trcon}), has already been determined through the solution of the plastic evolution problem described in section \ref{ch3:plasevo}. This force is evaluated based on $\Bsp\big(\bm{\varepsilon}^{(j)},\bm{\varepsilon}^{\mathrm{p}(j)},\bm{\varepsilon}^{\mathrm{r}(j)},\alpha^{(j-1)}\big)$, where $j$ refers to the current iteration counter. To rewrite the equation \eqref{c6:sd} in its volumetric and deviatoric parts, first $\bm{\mathsf{H}}^{\mathrm{kin}'}(\alpha)$ is derived as:
\begin{equation}
\bm{\mathsf{H}}^{\mathrm{kin}'}(\alpha)=\frac{g'(\alpha)\bm{\mathsf{C}}}{\big(1-g(\alpha)\big)^2}.
\label{c3:hkinprime}
\end{equation}
Using equations \eqref{c3:Cdam} and \eqref{c3:decomhkin}, together with degradation functions \eqref{c3:gk}, \eqref{c3:gmu}, and \eqref{c3:HkinKmu}, the damage driving force \eqref{c6:sd} is expressed in the following form
\begin{equation}
\begin{aligned}
s^\mathrm{d}(\bm{\varepsilon},\bm{\varepsilon}^\mathrm{p},\alpha)=\begin{dcases}
-\frac{1}{2}g_K'(\alpha)K\,(\tr\Be)^2-g_\mu'(\alpha)\mu\,\Be_\mathrm{dev}:\Be_\mathrm{dev}  &\quad \text{if } \tr{\Bsp}\big(\bm{\varepsilon},\bm{\varepsilon}^\mathrm{p},\bm{\varepsilon}^\mathrm{r},\alpha^{(j-1)}\big)=0,  \\
-\frac{g_K'(\alpha)K}{2[1-g_K(\alpha)]^2}\,(\tr\Bep)^2 - \frac{g_\mu'(\alpha)\mu}{[1-g_\mu(\alpha)]^2}\Bep_\mathrm{dev}:\Bep_\mathrm{dev} & \quad \text{if } \tr{\Bsp}\big(\bm{\varepsilon},\bm{\varepsilon}^\mathrm{p},\bm{\varepsilon}^\mathrm{r},\alpha^{(j-1)}\big)<0.
\end{dcases}
\end{aligned}
\label{c3:sdalg}
\end{equation}
To ensure that the irreversibility condition and equation \eqref{c3:weakd} are satisfied, the \emph{history field} approach is employed, which is originally proposed by Miehe et al. \cite{miehe2010b}. Therefore, the maximum time history values of the individual terms in the crack driving force are employed as follows
\begin{equation}
\begin{aligned}
\mathcal{H}_{K\mathrm{I}}(\bm{x},t)&\coloneqq \max_{s\in[0,t]}\frac{1}{2}K\big[\tr\Be(\bm{x},s)\big]^2, \\
\mathcal{H}_{\mu\mathrm{I}}(\bm{x},t)&\coloneqq\max_{s\in[0,t]}\mu\,\Be_\mathrm{dev}(\bm{x},s):\Be_\mathrm{dev}(\bm{x},s), \\
\mathcal{H}_{K\mathrm{II}}(\bm{x},t)&\coloneqq\max_{s\in[0,t]}\frac{1}{2}K\big[\tr\Bep(\bm{x},s)\big]^2,  \\
\mathcal{H}_{\mu\mathrm{II}}(\bm{x},t)&\coloneqq \max_{s\in[0,t]}\mu\,\Bep_\mathrm{dev}(\bm{x},s):\Bep_\mathrm{dev}(\bm{x},s),
\end{aligned}
\label{c3:histfield}
\end{equation}
and $s^\mathrm{d\,hist}$ is defined as
\begin{equation}
\begin{aligned}
s^\mathrm{d\,hist}(\bm{\varepsilon},\bm{\varepsilon}^\mathrm{p},\alpha)\coloneqq \begin{dcases}
-g_K'(\alpha)\mathcal{H}_{K\mathrm{I}}-g_\mu'(\alpha)\mathcal{H}_{\mu\mathrm{I}}& \quad \text{if } \tr{\Bsp}\big(\bm{\varepsilon},\bm{\varepsilon}^\mathrm{p},\bm{\varepsilon}^\mathrm{r},\alpha^{(j-1)}\big)=0,  \\
-\frac{g_K'(\alpha)}{[1-g_K(\alpha)]^2}\,\mathcal{H}_{K\mathrm{II}} - \frac{g_\mu'(\alpha)}{[1-g_\mu(\alpha)]^2}\mathcal{H}_{\mu\mathrm{II}} & \quad \text{if } \tr{\Bsp}\big(\bm{\varepsilon},\bm{\varepsilon}^\mathrm{p},\bm{\varepsilon}^\mathrm{r},\alpha^{(j-1)}\big)<0. 
\end{dcases}
\end{aligned}
\label{c3:sdalg2}
\end{equation}

Subsequently, equation \eqref{c3:weakd} is written as
\begin{equation}
\int_\Omega \bigg( -s^\mathrm{d\,hist}(\bm{\varepsilon},\bm{\varepsilon}^\mathrm{p},\alpha)\,\tilde{\alpha}  +h(\mathcal{F}) \frac{{G}_\mathrm{c}(\bm{s}^\mathrm{p})}{{\ell}}\big(\alpha\,\tilde{\alpha}+{\ell}^{2}\nabla\alpha\cdot\nabla\tilde{\alpha}\big) \bigg) \,\mathrm{d}\bm{x}  = 0 \quad \forall\,\tilde{\alpha}\in\mathrm{H}^1(\Omega;\mathbb{R}).
\label{c3:weakd2}
\end{equation}
Finally, a standard Newton-Raphson scheme is used to handle the non-linear degradation functions in the crack driving force \eqref{c3:sdalg2}, where the linearization of equation \eqref{c3:weakd2} is given by
\begin{equation}
\begin{aligned}
&\int_\Omega \bigg[\bigg(-\frac{\partial s^\mathrm{d\,hist}(\bm{\varepsilon},\bm{\varepsilon}^\mathrm{p},\alpha^{(k)})}{\partial\alpha^{(k)}} + h(\mathcal{F})\frac{{G}_\mathrm{c}(\bm{s}^\mathrm{p})}{{\ell}}\bigg)\big({\alpha}^{(k+1)}-{\alpha}^{(k)}\big)\tilde{\alpha}+ {\ell}\,h(\mathcal{F}){G}_\mathrm{c}(\bm{s}^\mathrm{p})\nabla\big[{\alpha}^{(k+1)}-{\alpha}^{(k)}\big]\cdot\nabla\tilde{\alpha}\bigg] \,\mathrm{d}\bm{x}\\ 
&=\int_\Omega \bigg[ s^\mathrm{d\,hist}(\bm{\varepsilon},\bm{\varepsilon}^\mathrm{p},\alpha^{(k)})\,\tilde{\alpha}  - h(\mathcal{F})\frac{{G}_\mathrm{c}(\bm{s}^\mathrm{p})}{{\ell}}\big(\alpha^{(k)}\,\tilde{\alpha}+{\ell}^{2}\nabla\alpha^{(k)}\cdot\nabla\tilde{\alpha}\big) \bigg] \,\mathrm{d}\bm{x}  \eqqcolon r_{\alpha}^{(k)} \quad \forall\,\tilde{\alpha}\in\mathrm{H}^1(\Omega;\mathbb{R}).
\end{aligned}
\label{c3:weakdlin}
\end{equation}
This equation is iteratively solved for $\alpha^{(k+1)}$ until the residual satisfies $\big\vert r_{\alpha}^{(k)}\big\vert\leq \mathtt{TOL}_{\alpha}$, where $\mathtt{TOL}_{\alpha}$ is a prescribed tolerance.

\subsection{Plastic evolution problem}\label{ch3:plasevo}
The plastic strain $\Bep$ and ratcheting strain $\Ber$ are computed by solving the plastic evolution equations while keeping the displacement $\Bu$ and damage variable $\alpha$ fixed. An implicit backward Euler scheme is used, with the non-associative flow rule discretized at each iteration as:
\begin{equation}
\Bep=\Bep_n + \Delta \Bep \quad \text{with} \quad \Delta\Bep=\Delta\gamma\,\hat{\bm{n}}, \quad \hat{\bm{n}}\in\partial g^\mathrm{p}(\Bsp).
\label{c6:flo}
\end{equation}
For simplicity, the iteration counter $k$ is omitted. The incremental approximation of the plastic multiplier is denoted as $\Delta\gamma$, with $\dot{\gamma}\coloneqq\lambda$. The trial stresses are computed as:
\begin{equation}
\bm{s}^\mathrm{p\,trial}\coloneqq\Bsig^\mathrm{trial} -\bm{\mathsf{H}}^\mathrm{kin}(\alpha)\Bep_n \quad \text{with} \quad \Bsig^\mathrm{trial}\coloneqq \bm{\mathsf{C}}:(\Be-\Bep_n-\Ber_n).
\label{c6:sptr}
\end{equation}
The resulting incremental system is:
\begin{equation}
\begin{dcases}\Bep=\Bep_n + \Delta\gamma\,\hat{\bm{n}}, \quad \hat{\bm{n}}\in\partial g^\mathrm{p}(\Bsp), \\
\Bsp=\bm{s}^\mathrm{p\,trial}-\Delta\gamma\,\big[\bm{\mathsf{C}}+\bm{\mathsf{H}}^\mathrm{kin}(\alpha)\big]:\hat{\bm{n}},\\
f(\Bsp)\leq 0, \quad  \Delta\gamma\geq 0, \quad \Delta\gamma \, f(\Bsp) = 0. \end{dcases}
\label{c6:psys}
\end{equation}
The elastic predictor/plastic corrector algorithm is used to solve the aforementioned system \cite{de2011a}, for which the yield function is first evaluated for the trial state. If $ f(\bm{s}^\mathrm{p\,trial}) \leq 0$, then the trial state is admissible and the solution to \eqref{c6:psys} takes the following form 
\begin{equation}
\begin{dcases}
\Delta\gamma = 0,\\ \Bep=\Bep_n, \\ \Ber=\Ber_n, \\
\Bsp=\bm{s}^\mathrm{p\,trial}, \end{dcases} \quad \implies \quad \begin{dcases}\Bsig=\Bsig^\mathrm{trial}, \\ \bm{\mathsf{C}}^{\mathrm{ep}}=\frac{\partial\Bsig}{\partial\Be}=\bm{\mathsf{C}}.\end{dcases}
\label{c6:psyssolu}
\end{equation}
If $f(\bm{s}^\mathrm{p\,trial})>0$, then a corrector step should be applied to return the trial state back to the set of admissible stresses and hence, the incremental system (equation \eqref{c6:psys}) is given by
\begin{equation}
\begin{dcases}\Bep=\Bep_n + \Delta\gamma\,\hat{\bm{n}}, \quad \hat{\bm{n}}\in\partial g^\mathrm{p}(\Bsp), \\
\Bsp=\bm{s}^\mathrm{p\,trial}-\Delta\gamma\,\big[\bm{\mathsf{C}}+\bm{\mathsf{H}}^\mathrm{kin}(\alpha)\big]:\hat{\bm{n}},\\
f(\Bsp)= 0, \end{dcases}
\label{c6:psys_2}
\end{equation}
Now it should be determined if $\Bsp$ is located on the smooth part of the Drucker-Prager cone or at the apex. For this purpose, an implicit return-mapping scheme similar to Sysala et al.~\cite{sysala2016a} is employed. First, the generalized stress $\Bsp$ is decomposed into deviatoric and volumetric parts:
\begin{equation}
\begin{aligned}
\Bsp_\mathrm{dev}&\equiv\bm{s}^\mathrm{p\,trial}_\mathrm{dev}-\Delta\gamma\,\big[2\mu+H^\mathrm{kin}_\mu(\alpha)\big]\hat{\bm{n}}_\mathrm{dev} \\ \text{and} \quad \tr\Bsp&\equiv \tr\bm{s}^\mathrm{p\,trial} -3\sqrt{6}\Delta\gamma\,A_\theta\big[K+H^\mathrm{kin}_K(\alpha)\big].
\end{aligned}
\label{c6:sp_sptr}
\end{equation}
Subsequently, the deviatoric part of the trial stress $\bm{s}^\mathrm{p\,trial}_\mathrm{dev}$ is expressed as
\begin{equation}
\bm{s}^\mathrm{p\,trial}_\mathrm{dev} =\begin{dcases} \bigg(1 + \dfrac{\Delta\gamma\,\big[2\mu+H^\mathrm{kin}_\mu (\alpha)\big]}{\Vert\Bsp_\mathrm{dev} \Vert} \bigg) \Bsp_\mathrm{dev}  & \text{if} \quad \Vert \Bsp_\mathrm{dev} \Vert > 0, \\
\Delta\gamma\,\big[2\mu+H^\mathrm{kin}_\mu(\alpha)\big]\hat{\bm{n}}_\mathrm{dev}, \quad \Vert\hat{\bm{n}}_\mathrm{dev}\Vert \leq 1 & \text{if} \quad \Vert \Bsp_\mathrm{dev} \Vert = 0.
\end{dcases}
\label{c6:sptr2}
\end{equation}
The corresponding norm is defined as
\begin{equation}
\Vert \Bsp_\mathrm{dev} \Vert = \Big\langle \Vert\bm{s}^\mathrm{p\,trial}_\mathrm{dev} \Vert - \Delta\gamma\, \big[2\mu+H^\mathrm{kin}_\mu (\alpha)\big]\Big\rangle_+\,,
\label{c6:sp_ramp}
\end{equation}
where $\langle\Box\rangle_+\coloneqq(\Box+\vert\Box\vert)/2$. To express the yield condition in terms of the trial stress and the plastic multiplier, an auxiliary function is defined as:
\begin{equation}
\begin{aligned}
Q(\Delta\gamma)&\coloneqq\Big< \Vert\bm{s}^\mathrm{p\,trial}_\mathrm{dev}\Vert - \Delta\gamma\,\big[2\mu+H^\mathrm{kin}_\mu (\alpha)\big] \Big>_+ \\&+\sqrt{\frac{2}{3}}A_\varphi\Big( \tr\bm{s}^\mathrm{p\,trial} -3\sqrt{6}\,\Delta\gamma\,A_\theta\big[K+H^\mathrm{kin}_K(\alpha) \big] \Big).
\end{aligned}
\label{c6:q}
\end{equation}
This involves replacing equation \eqref{c6:sp_sptr} in equation \eqref{c3:fp}. To determine whether $\Bsp$ lies on the smooth region or at the apex of the yield surface, the following procedure is applied:
\begin{enumerate}
\item If $Q\big({\Vert\bm{s}^\mathrm{p\,trial}_\mathrm{dev} \Vert}/{\big[2\mu+H^\mathrm{kin}_\mu (\alpha)\big]}\big)<0$, then $\Bsp$ lies on the smooth part of the yield surface. The solution is then given by
\begin{equation}
\begin{dcases}
\Delta\gamma = \dfrac{ \Vert\bm{s}^\mathrm{p\,trial}_\mathrm{dev} \Vert+\sqrt{2/3}\,A_\varphi\tr\bm{s}^\mathrm{p\,trial}}{2\mu+H^\mathrm{kin}_\mu (\alpha)+6A_\varphi A_\theta\big[ K+H^\mathrm{kin}_K(\alpha) \big]},\\ 
\Bep=\Bep_n + \Delta\gamma\,(\hat{\bm{n}}_\mathrm{dev}^\mathrm{trial}+\sqrt{2/3}\,A_\theta \bm{1} ),\\
\Bsp=\bm{s}^\mathrm{p\,trial}- \Delta\gamma\,\big[2\mu+H^\mathrm{kin}_\mu (\alpha)\big]\hat{\bm{n}}_\mathrm{dev}^\mathrm{trial} - \sqrt{6}\,\Delta\gamma\,A_\theta\big[K+H^\mathrm{kin}_K(\alpha) \big] \bm{1}. \end{dcases}
\label{c6:finalsys}
\end{equation}
The stress tensor is updated as
\begin{equation}
\begin{aligned}
\Bsig&=\Bsig^\mathrm{trial}-\Delta\gamma\,\bm{\mathsf{C}}:(\hat{\bm{n}}_\mathrm{dev}^\mathrm{trial}+\sqrt{2/3}\,A_\theta \bm{1})\\ &=\Bsig^\mathrm{trial}-\Delta\gamma\,\big(2\mu\,\hat{\bm{n}}_\mathrm{dev}^\mathrm{trial}+\sqrt{6}\,K A_\theta\bm{1}\big),
\end{aligned}
\label{c6:sigmaup}
\end{equation} 
and the consistent tangent becomes
\begin{equation}
\begin{aligned}
\bm{\mathsf{C}}^\mathrm{ep}=\frac{\partial\Bsig}{\partial\Be}&=\bm{\mathsf{C}}-\big(2\mu\,\hat{\bm{n}}_\mathrm{dev}^\mathrm{trial}+\sqrt{6}\,K A_\theta\bm{1}\big)\otimes\frac{\partial(\Delta\gamma)}{\partial\Be}-2\mu\,\Delta\gamma\frac{\partial\hat{\bm{n}}_\mathrm{dev}^\mathrm{trial}}{\partial\Be} \\
&=\bm{\mathsf{C}}-\frac{\big(2\mu\,\hat{\bm{n}}_\mathrm{dev}^\mathrm{trial}+\sqrt{6}\,K A_\theta\bm{1}\big) \otimes \big(2\mu\,\hat{\bm{n}}_\mathrm{dev}^\mathrm{trial}+\sqrt{6}\,K A_\varphi\bm{1}\big)}{[2\mu+H^\mathrm{kin}_\mu (\alpha)]+6A_\varphi A_\theta\big[ K+H^\mathrm{kin}_K(\alpha) \big]} \\
&\hspace{2.2cm}-\frac{4\mu^2\Delta\gamma}{\Vert\bm{s}^\mathrm{p\,trial}_\mathrm{dev} \Vert}\bigg(\bm{\mathsf{I}}-\frac{1}{3}\bm{1}\otimes\bm{1}-\hat{\bm{n}}_\mathrm{dev}^\mathrm{trial}\otimes\hat{\bm{n}}_\mathrm{dev}^\mathrm{trial}\bigg).
\end{aligned}
\label{c6:tanstif}
\end{equation}

The ratcheting strain is now obtained by
\begin{equation}
\begin{aligned}
\Ber&=\Ber_n + \Delta\gamma\,\hat{\bm{r}}, \\
&=\Ber_n + \Delta\gamma\,(\beta_{\mathrm{\mu}} \hat{\bm{r}}_{\mathrm{dev}}+\sqrt{2/3} \,\beta_{\mathrm{K}}A_\theta), \quad \hat{\bm{r}}_{\mathrm{dev}}=\dfrac{\Bsr_\mathrm{dev}}{\Vert\Bsr_\mathrm{dev}\Vert}=\dfrac{\Bsig_\mathrm{dev}}{\Vert \Bsig_\mathrm{dev}\Vert}.
\end{aligned}
\label{rstrain}
\end{equation}

\item If $Q\big({\Vert\bm{s}^\mathrm{p\,trial}_\mathrm{dev} \Vert}/{\big[2\mu+H^\mathrm{kin}_\mu (\alpha)\big]}\big)\geq0$, then $\Bsp$ lies at the apex of the cone, corresponding to open microcracks with no frictional sliding. The stress tensor is given by
\begin{equation}
\Bsig=\bm{\mathsf{C}}^\mathrm{dam}(\alpha):\bm{\varepsilon},
\label{c3:sigmaup2}
\end{equation}
and the consistent tangent is
\begin{equation}
\bm{\mathsf{C}}^\mathrm{ep}=\frac{\partial\Bsig}{\partial\Be}=\bm{\mathsf{C}}^\mathrm{dam}(\alpha).
\label{c3:tanstif2}
\end{equation}
\end{enumerate}
At each iteration, these expressions are used to compute $\Bsig$ and $\bm{\mathsf{C}}^\mathrm{ep}$, which are required for solving the equilibrium equation in \eqref{c3:weakualg}.

\section{Numerical simulations}\label{ch6:numsim}
This section presents numerical simulations to highlight the key features of the micromechanics-based phase-field model \cite{ulloa2022a}, its extension to fatigue \cite{sarem2025a}, and the proposed incorporation of ratcheting-driven cyclic plasticity. First, a three-point bending test under cyclic loading is investigated to evaluate the model’s ability to reproduce fatigue behavior through comparison with experimental results. Second, high-cycle fatigue performance is assessed using a single element, from which an $S$–$N$ curve is obtained based on two different fatigue degradation functions. Third, the influence of ratcheting is explored through material point simulations under cyclic axial compression and tension–compression loading. Finally, a square specimen with a central hole is analyzed under cyclic force-controlled loading to demonstrate the model’s capability in low-cycle fatigue scenarios where plasticity plays an important role.

\subsection{Calibration of fatigue response using a three-point bending test \label{ch3:tpb}}
\begin{figure}[htb!]
  \centering
    \includegraphics[scale=1.2]{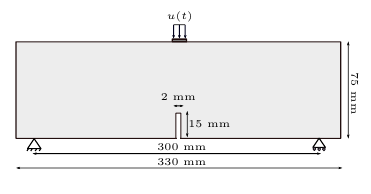}
\caption{Boundary conditions of a centrally notched concrete beam subjected to three-point bending.}
\label{fig3:tpb}
\end{figure}

The cyclic response of the micromechanics-based phase-field model is evaluated through simulation of a centrally notched concrete beam in a three-point bending test, following the experimental study of Keerthana and Kishen \cite{keerthana2018a}. A 2D finite element analysis is performed under the plane strain assumption. The concrete beam is placed on two supports, and a concentrated load is applied at the midpoint. The geometry and boundary conditions are shown in figure~\ref{fig3:tpb}.

Material parameters are adopted from the literature \cite{keerthana2018a}, with Poisson’s ratio $\nu = 0.12$, Young’s modulus $E=35305$ MPa, mode I fracture energy $G_\mathrm{cI} = 0.112$ $\frac{\mathrm{N}}{\mathrm{mm}}$, mode II fracture energy $G_\mathrm{cII} = 0.112$ $\frac{\mathrm{N}}{\mathrm{mm}}$ and friction and dilation angles of $\varphi=48^\circ$ and $\theta=30^\circ$, respectively. The beam is subjected to cyclic compressive loading up to a maximum force of 2.25 kN. A minimum element size of $h=2$ mm is used in the region where cracking is expected, resulting in a total of 3353 elements and 3478 nodes. The length scale is set to $l = 5$ mm, satisfying the criterion $\frac{l}{h} \geq 2$.

\begin{figure}[htb!]
\centering \includegraphics[scale=0.45]{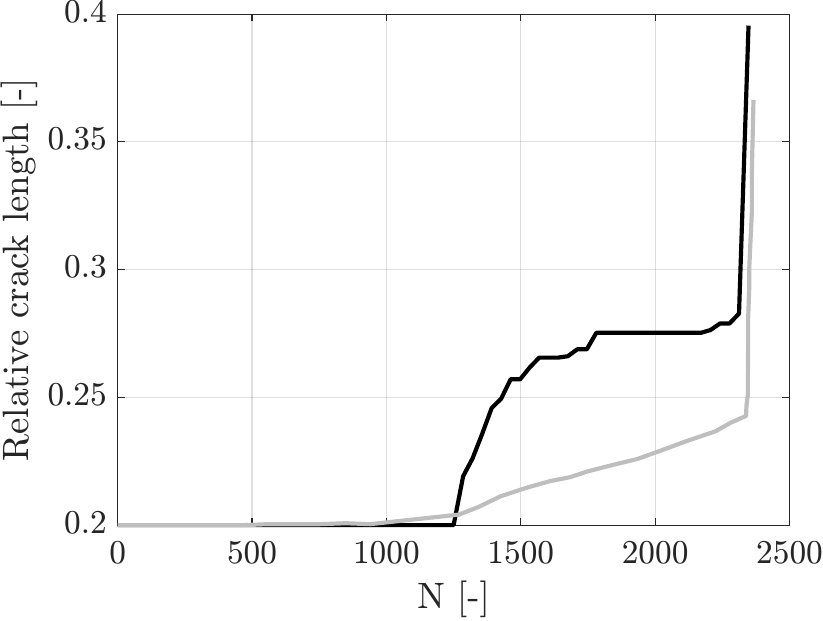}
\caption{Evolution of the relative crack length with the number of cycles in the three-point bending beam for the experiment ($\textcolor{greylight}{\full}$) and the simulation ($\textcolor{black}{\full}$).}
\label{fig5:life}
\end{figure}

Fatigue effects are incorporated through the logarithmic degradation function (equation~\eqref{c5:log}). The fatigue threshold, $\mathcal{F}_0=1\times10^{-4}\,\frac{\mathrm{N}}{\mathrm{mm}^2}$, and the fatigue parameter, $k=1$, are calibrated such that crack initiation occurs after approximately 1250 loading cycles, consistent with the experimental observations. The fatigue response of the three-point bending test is presented in figure~\ref{fig5:life}, where the relative crack length, $\frac{a}{d}$, is plotted against the number of loading cycles, with $a$ denoting the crack length and $d$ the specimen depth. After crack initiation, the simulation captures a regime of stable crack propagation followed by sudden unstable failure after a total of 2347 cycles.

The predictive capability of the model is assessed by comparing the simulated crack-growth evolution with the experimental measurements reported by~\citep{keerthana2018a}, without further adjustment of the fatigue parameters after calibration of the crack-initiation point. The model predicts final failure after 2347 cycles, compared with 2366 cycles observed experimentally, corresponding to a difference of approximately 0.8\%. This good agreement demonstrates the ability of the proposed framework to accurately predict the overall fatigue life under cyclic loading.

While the predicted fatigue life agrees remarkably well with the experimental observations, differences are observed in the detailed crack-growth evolution. The experimental results exhibit an approximately constant crack-growth rate over a large portion of the fatigue life, whereas the numerical model predicts an initial rapid crack growth, followed by a relatively stable propagation phase and an accelerated growth stage preceding final failure. The latter behavior arises from the progressive accumulation of fatigue degradation within the phase-field framework, which eventually leads to unstable crack growth once a critical state is reached.

It is worth noting that the three-stage crack-growth behavior predicted by the present numerical simulations is consistent with similar trends reported in experimental studies on quasi-brittle materials \cite{baktheer2021a, chen2025a}. The experimental dataset of \citet{keerthana2018a} was selected because it provides a comprehensive set of material properties and testing parameters required for a direct numerical comparison, whereas many available fatigue studies do not report sufficient information for a consistent calibration and validation procedure.

The difference between the measured and simulated crack-growth histories may be attributed to factors that are not explicitly represented in the present model, including material heterogeneity and spatial variability of material properties. In addition, uncertainties associated with experimental crack-length measurements may affect the reported crack-growth evolution. Nevertheless, the model successfully captures the onset of crack propagation, the overall fatigue-damage evolution, and the final fatigue life, indicating that the dominant mechanisms governing fatigue failure are adequately represented.

\begin{figure} [htb!]
\centering \includegraphics[scale=1]{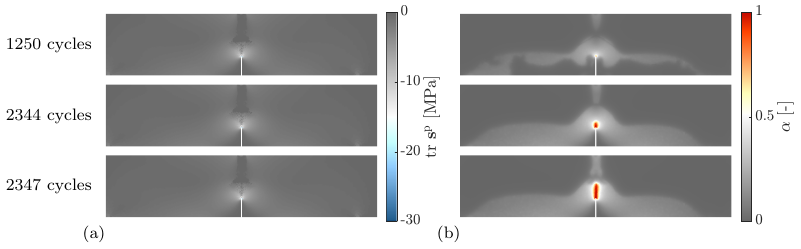}
\caption{Fracture process at different loading cycles for the three-point bending beam, showing (a) the hydrostatic generalized stress $\tr{\Bsp}$ and (b) the corresponding damage profiles.}
\label{fig5:tpb_damtr}
\end{figure}

To further illustrate the cyclic response of the notched beam, figure \ref{fig5:tpb_damtr}(a) shows contours of the hydrostatic generalized stress $\tr\bm{s}^\mathrm{p}$ at selected loading cycles. Regions where $\tr\bm{s}^\mathrm{p}$ vanishes indicate the progression of tensile crack growth, consistent with the characteristic response observed in three-point bending tests, which are commonly used to evaluate the tensile strength of materials \cite{you2020a}. Figure \ref{fig5:tpb_damtr}(b) illustrates the damage contours across different cycles. Crack initiation is observed at the notch tip after 1250 cycles. As loading progresses, the crack propagates toward the midspan of the beam. The comparison of damage profiles at cycles 2344 and 2347 clearly shows an abrupt failure phase, highlighting the sudden transition from stable fatigue growth to unstable crack propagation. This example demonstrates the capability of the proposed model to capture tensile fatigue crack evolution and final failure under cyclic loading conditions.

%

\subsection{High-cycle fatigue of a single 2D element}\label{ch5:highfat}
To investigate the model’s response under high-cycle loading, a simplified numerical test is performed on a single 2D element. This approach reduces the computational cost associated with large boundary value problems involving thousands of elements and degrees of freedom over a huge number of time steps, while still providing valuable insight into the underlying mechanisms. The applied biaxial strain is varied from 0.8 to 10, corresponding to a range of cycles from $10^8$ down to 1 at failure. Failure for this material point is defined as the point at which the damage parameter reaches $\alpha = 0.97$.

The material properties adopted for this simulation are: Poisson's ratio $\nu=0.2$, Young's modulus $E=1$ MPa, degradation constant $b=1$, mode I fracture energy $G_\mathrm{cI}=10$ $\frac{\mathrm{N}}{\mathrm{mm}}$, and mode II fracture energy $G_\mathrm{cII}=100$ $\frac{\mathrm{N}}{\mathrm{mm}}$. The friction and dilation coefficients are set to $A_\varphi=0.1$ and $A_\theta=0.075$, respectively. Fatigue effects are included using the logarithmic degradation function (equation \eqref{c5:asymp}), with a threshold $\mathcal{F}_0=7$ $\frac{\mathrm{N}}{\mathrm{{mm}^2}}$ and fatigue parameter $k=0.7$.

\begin{figure}[htb!]
\centering \includegraphics[scale=0.9]{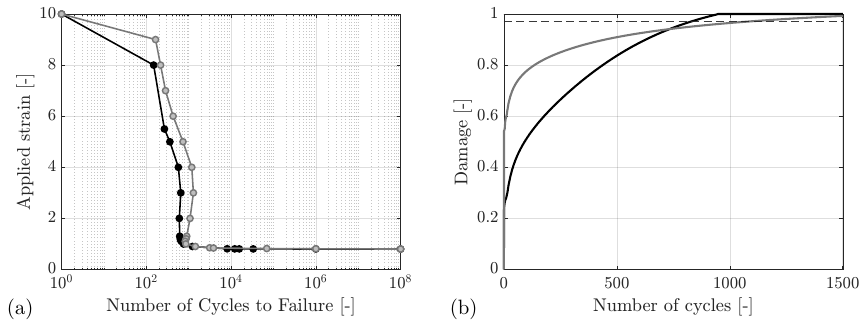}
\caption{Results of the single element under high-cycle loading: (a) $S-N$ curve showing the applied strain over number of cycles for the asymptotic degradation function ($\full$) and the logarithmic function ($\textcolor{greydark}{\full}$); (b) damage evolution for the logarithmic function with applied strains $\varepsilon_{xx}=\varepsilon_{yy}=1$ ($\full$) and $\varepsilon_{xx}=\varepsilon_{yy}=2$ ($\textcolor{greydark}{\full}$).}
\label{fig5:sn_homo}
\end{figure}

The resulting $S-N$ curve is shown in figure \ref{fig5:sn_homo}(a). Also known as the W\"ohler curve, it illustrates how a material fails under cyclic loading \cite{schreiber2020a}, relating the maximum number of cycles a material can endure before failure to the constant applied load (or strain, as in this example). The curve captures several characteristic features of fatigue, including the influence of the load amplitude, the presence of an upper stress amplitude corresponding to the material’s monotonic strength, and the possible existence of a fatigue threshold \cite{carrara2020a}.

Results are presented for two fatigue degradation functions introduced earlier: the asymptotic and the logarithmic forms (equations \eqref{c5:asymp} and \eqref{c5:log}). For the logarithmic function, an unexpected increase in the number of cycles was observed with increasing applied strain. To investigate this behavior, damage profiles for two applied strain levels are shown in figure \ref{fig5:sn_homo}(b). The evolution of damage confirms that higher applied strains lead to faster damage growth. However, due to the specific form of the logarithmic degradation function, the growth rate decreases at high damage levels, resulting in lower overall damage compared to the lower strain case.

This analysis indicates that the unusual number of cycles observed for certain applied strains in the logarithmic function is a result of the mathematical formulation of the degradation function, rather than an intrinsic material property. These findings emphasize the importance of performing sensitivity analyses on fatigue properties during model calibration.

%

\subsection{Ratcheting response of a material point}\label{ch6:homo}
To assess the effect of the ratcheting mechanism, the response of a material point under cyclic loading is analyzed. To this end, the following properties are considered: Poisson's ratio $\nu=0.2$, Young's modulus $E=1$ MPa, degradation constant $b=2$, initial damage $\alpha_0=1\times10^{-4}$, mode I fracture energy $G_\mathrm{cI}=5$ $\frac{\mathrm{N}}{\mathrm{mm}}$, mode II fracture energy $G_\mathrm{cII}=6$ $\frac{\mathrm{N}}{\mathrm{mm}}$ and friction and dilation coefficients $A_{\varphi}=0.1$ and $A_{\theta}=0.075$, respectively. It should be noted that these parameters are not physically realistic values, but are chosen for illustrative purposes.


In the first example, the material point is subjected to 20 load cycles under both stress-controlled and strain-controlled conditions. In the stress-controlled case, the applied axial stress ranges from 0 to $-3$ MPa, while in the strain-controlled case, the strain varies between 0 and $-4$. To focus only on the ratcheting behavior, fatigue degradation is disabled by setting the degradation threshold to $\mathcal{F}_0 = \infty$. The ratcheting parameters are set to $\beta_K = 0.2$ and $\beta_\mu = 0.1$ in this example.

\begin{figure} [htb!]
\centering \includegraphics[scale=1]{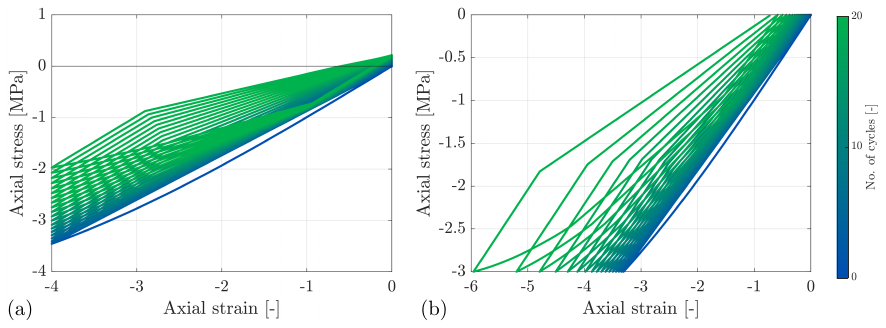}
\caption{Axial stress $\sigma_{xx}$ versus axial strain $\varepsilon_{xx}$ response of the material point under (a) cyclic strain-controlled loading and (b) cyclic stress-controlled loading.}
\label{fig6:homofd}
\end{figure}

Figure \ref{fig6:homofd} shows the stress–strain response of the material point under (a) strain-controlled and (b) stress-controlled loading. In the strain-controlled case, the applied strain follows a prescribed cyclic pattern, while the stress evolves in response. The material exhibits cyclic softening, commonly referred to as stress relaxation. With continued loading, the stress amplitude gradually stabilizes, indicating the accumulation of permanent deformation without the onset of instability. In contrast, the stress-controlled case involves cyclic variation of the applied stress, allowing the strain to evolve freely. This leads to ratcheting behavior, where the strain progressively accumulates over cycles. As the number of cycles increases, the strain amplitude grows significantly, showing the unbounded nature of strain ratcheting. This highlights the importance of ratcheting under stress-controlled conditions, where the absence of strain constraints allows plastic deformation to accumulate. Over time, this increase in plastic strain can lead to instability or even failure, especially in low-cycle fatigue situations.

\begin{figure} [htb!]
\centering \includegraphics[scale=1]{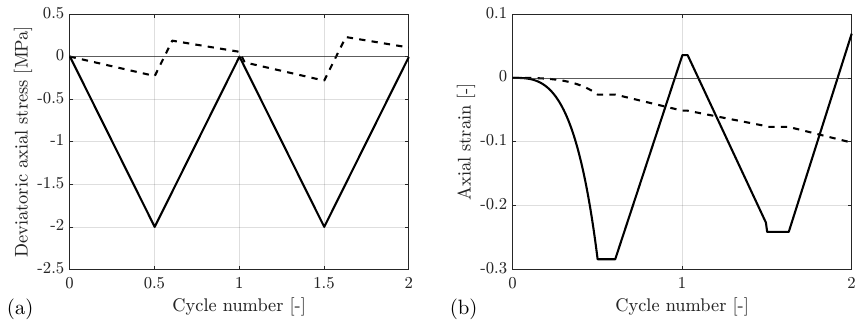}
\caption{Response of the material point under axial stress-controlled loading during the first two cycles: (a) evolution of the deviatoric stress component $\sigma_{\mathrm{dev} \, xx}$ ($\full$) and the deviatoric generalized stress component $\Bsp_{\mathrm{dev} \, xx}$ ($\dashed$); (b) time evolution of the axial plastic strain $\Bep_{xx}$ ($\full$) and the axial ratcheting strain $\Ber_{xx}$ ($\dashed$).}
\label{fig6:homos}
\end{figure}

Furthermore, the evolution of the response of the material point under stress-controlled loading is shown in figure \ref{fig6:homos}. In Figure \ref{fig6:homos}(a), the deviatoric stress $\sigma_{\mathrm{dev} , xx}$ and the deviatoric generalized stress $\Bsp_{\mathrm{dev} , xx}$ are plotted over the first two cycles. Figure \ref{fig6:homos}(b) presents the corresponding axial plastic strain $\Bep_{xx}$ and ratcheting strain $\Ber_{xx}$.

In the second example of the material point, uniaxial cyclic loading is applied to study the combined effects of ratcheting and fatigue. The loading is prescribed either in the form of stress cycles ranging from $-0.4$ MPa to 2 MPa over 20 cycles (stress-controlled case), or as strain cycles varying between $-0.4$ to $2$ over 40 cycles (strain-controlled case). To activate fatigue-induced degradation, the fatigue threshold is set to $\mathcal{F}_0=3$ $\mathrm{\frac{N}{{mm}^2}}$, and the logarithmic-type degradation function (equation \eqref{c5:log}) with $k = 0.1$ is employed.

\begin{figure} [htb!]
\centering \includegraphics[scale=1]{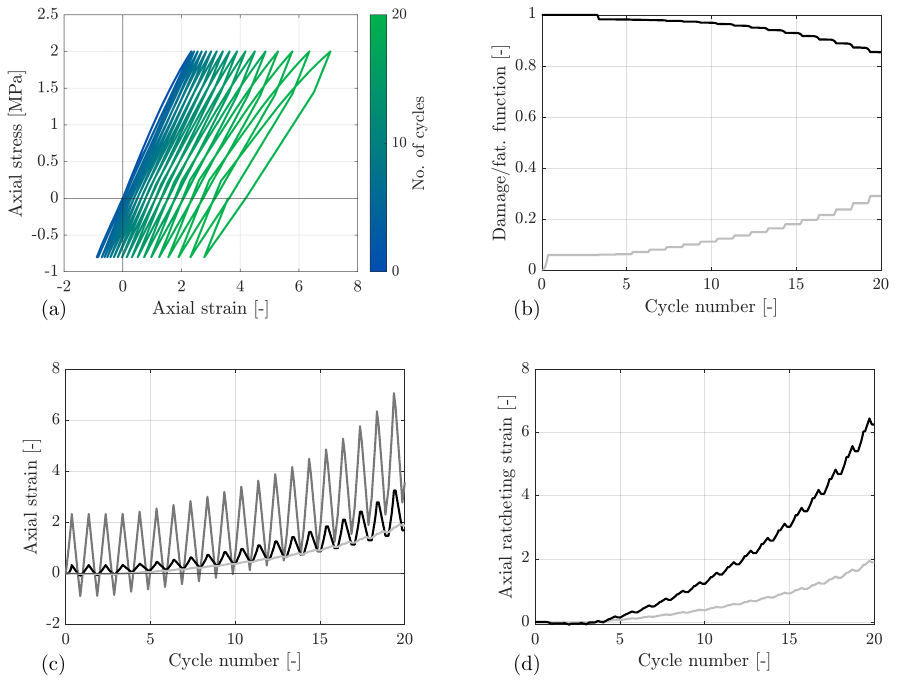}
\caption{Response of the material point under axial stress-controlled loading: (a) stress-strain curve, (b) evolution of damage ($\textcolor{greylight}{\full}$) and fatigue degradation function ($\full$), (c) evolution of axial total strain $\varepsilon_{xx}$ ($\textcolor{greydark}{\full}$), plastic strain $\Bep_{xx}$ ($\full$), and ratcheting strain $\Ber_{xx}$ ($\textcolor{greylight}{\full}$), and (d) evolution of axial ratcheting strain $\Ber_{xx}$ for two parameter sets: $\beta_K=0.2$, $\beta_\mu=0.1$ ($\textcolor{greylight}{\full}$) and $\beta_K=0.5$, $\beta_\mu =0.4$ ($\full$).}
\label{fig6:fatratch}
\end{figure}

Figure \ref{fig6:fatratch} shows the material response under stress-controlled loading. In figure \ref{fig6:fatratch}(a), the stress-strain curve reveals a clear degradation of stiffness due to damage and a progressive shift of the loop caused by ratcheting. The accumulation of strain and degradation of stiffness lead to accelerated cyclic softening. Figure \ref{fig6:fatratch}(b) shows the evolution of the damage variable and the fatigue degradation function. Damage initiates early due to the AT-2 formulation and increases gradually over cycles. Fatigue degradation begins after 3 cycles, reducing the fracture toughness and thus further accelerating the material's softening response. The evolution of the total strain, plastic strain, and ratcheting strain during loading cycles is shown in figure \ref{fig6:fatratch}(c). Since the stress is controlled, the strain is free to evolve. All strain measures grow over cycles, both in their amplitude and average magnitude, highlighting the cumulative inelastic deformation. Figure \ref{fig6:fatratch}(d) compares the evolution of the ratcheting strain for two different ratcheting parameter sets. As expected, the case with higher ratcheting parameters results in greater strain accumulation, confirming the model's sensitivity to these parameters.

\begin{figure} [htb!]
\centering \includegraphics[scale=1]{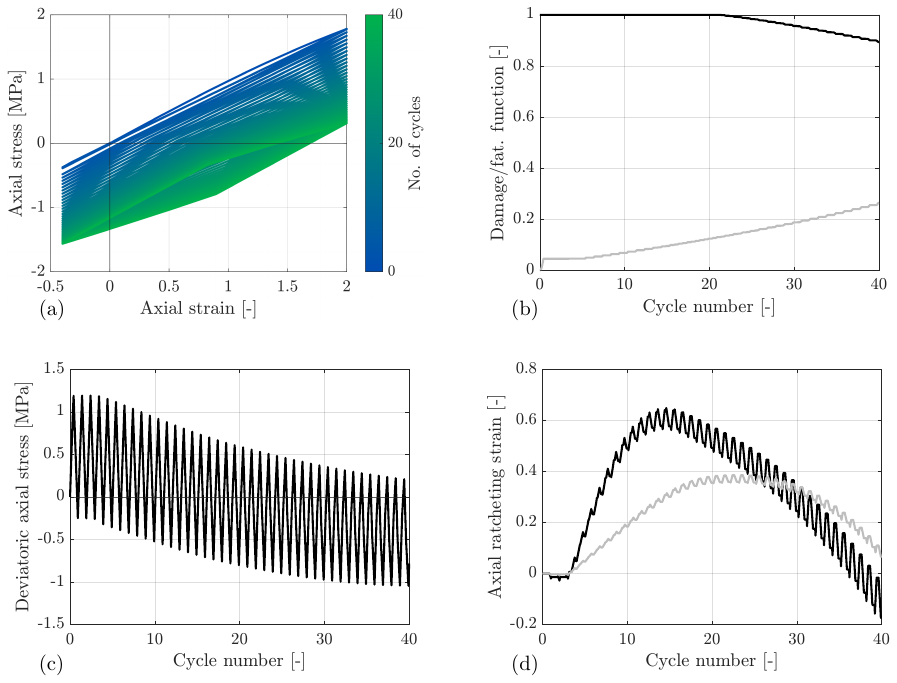}
\caption{Response of the material point under axial strain-controlled loading: (a) stress-strain curve, (b) evolution of damage ($\textcolor{greylight}{\full}$) and fatigue degradation function ($\full$), (c) evolution of the deviatoric stress component $\sigma_{\mathrm{dev} \, xx}$, and (d) evolution of axial ratcheting strain $\Ber_{xx}$ for two parameter sets: $\beta_K=0.2$, $\beta_\mu=0.1$ ($\textcolor{greylight}{\full}$), and $\beta_K=0.5$, $\beta_\mu =0.4$ ($\full$).}
\label{fig6:disfatratch}
\end{figure}

Figure \ref{fig6:disfatratch} shows the response of the material point under strain-controlled cyclic loading. In figure \ref{fig6:disfatratch}(a), the stress-strain response displays cyclic softening, observed as a reduction in stress amplitude with increasing cycles. The accumulation of inelastic deformation leads to a narrowing of the hysteresis loops, while the imposed strain range remains fixed. The shifting of the stress response toward lower values indicates the progressive damage, stress relaxation and ratcheting effects. The evolution of the damage variable and fatigue degradation function is shown in figure \ref{fig6:disfatratch}(b). Damage is activated from the beginning as a consequence of the AT-2 formulation, and the fatigue degradation function begins to decrease after 22 cycles, accelerating the loss of stiffness. Figure \ref{fig6:disfatratch}(c) shows the evolution of the deviatoric axial stress $\sigma_{\mathrm{dev},xx}$. The cycle-averaged (mean) stress gradually decreases due to cyclic softening induced by damage and fatigue degradation, reflecting stress relaxation under strain-controlled loading. The axial ratcheting strain (grey curve in figure \ref{fig6:disfatratch}(d)) increases while the mean stress is positive. After 14 cycles, when the mean stress becomes negative, the ratcheting strain begins to decrease. Figure \ref{fig6:disfatratch}(d) also compares $\Ber_{xx}$ for two sets of ratcheting parameters. In both cases the ratcheting strain accumulates over the cycles, and larger values of  $\beta_K$ and $\beta_\mu$ result in greater accumulation, confirming the model’s sensitivity to these parameters.

%

\subsection{Cyclic response of a perforated specimen}\label{ch6:perfs}
\begin{figure}[htb!]
\centering \includegraphics[scale=0.9]{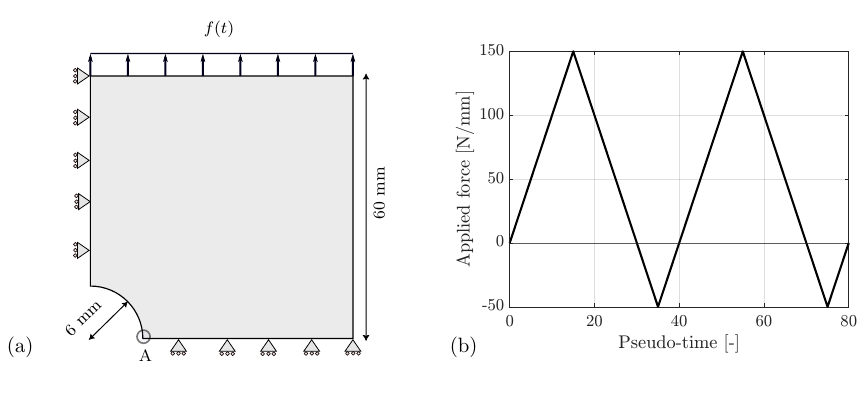}
\caption{(a) Boundary conditions applied to the perforated specimen and (b) imposed cyclic force over two loading cycles.}
\label{fig6:perf_bc}
\end{figure}

In this example, a square specimen with a central hole is analyzed under plane strain conditions and subjected to force-controlled loading. Due to symmetry conditions, only the top-right quarter of the specimen is modeled. A uniform load is applied along the top edge, and a regular mesh consisting of 800 elements is used. The geometry and boundary conditions are illustrated in figure \ref{fig6:perf_bc}(a). The applied cyclic force, varying from tension to compression, is shown over two loading cycles in figure \ref{fig6:perf_bc}(b). This test evaluates the model's capability to capture cyclic behavior involving both fatigue and ratcheting. A similar setup was previously investigated by Ulloa et al. \cite{ulloa2021a, ulloa2021b}, but without damage and fatigue effects.

The material properties adopted for this simulation are: Poisson's ratio $\nu=0.3$, Young's modulus $E=205$ GPa, degradation constant $b=1$, mode I fracture energy $G_\mathrm{cI}=10$ $\frac{\mathrm{N}}{\mathrm{mm}}$, mode II fracture energy $G_\mathrm{cII}=50$ $\frac{\mathrm{N}}{\mathrm{mm}}$ and length scale $l=2$ mm. The element size is chosen as $h=0.6$ mm, resulting in a total of 800 elements and 861 nodes. The friction and dilation coefficients are set to $A_\varphi=0.1$ and $A_\theta=0.075$, respectively. Fatigue effects are included using the asymptotic degradation function (equation \eqref{c5:asymp}), with a threshold of $\mathcal{F}_0=5$ $\frac{\mathrm{N}}{\mathrm{{mm}^2}}$. The ratcheting behavior is characterized by bulk and shear ratcheting parameters of $\beta_K=0.2$ and $\beta_\mu=0.1$, respectively.

\begin{figure}[htb!]
\centering \includegraphics[scale=0.5]{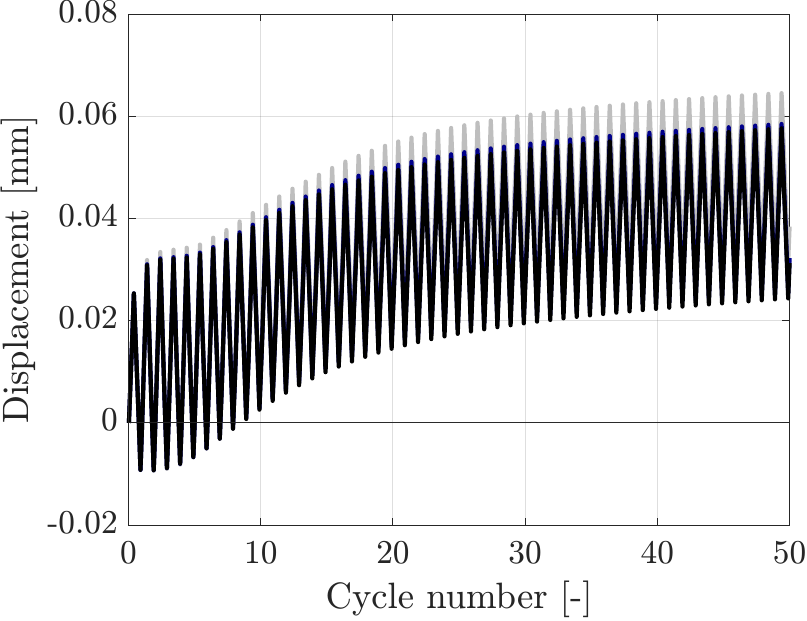}
\caption{Time history of the displacement for the perforated specimen under force loading computed with different mesh densities: 300 elements ($\textcolor{greylight}{\full}$), 800 elements ($\textcolor{blue}{\full}$), and 3500 elements ($\textcolor{black}{\full}$).}
\label{fig6:mesh}
\end{figure}

\begin{figure}[htb!]
\centering \includegraphics[scale=0.5]{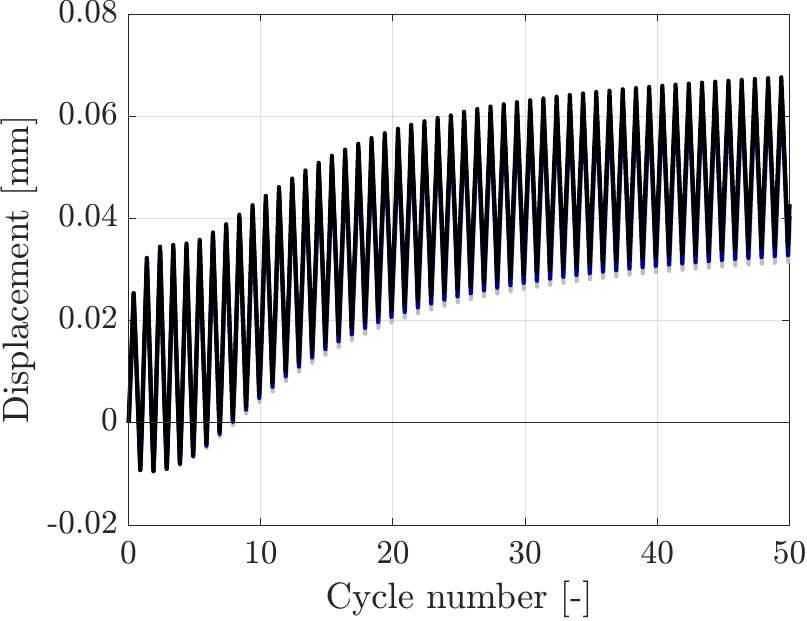}
\caption{Time history of the displacement for the perforated specimen under force loading computed with different time steps: $\Delta f=$ 10 $\frac{N}{mm.s}$ ($\textcolor{greylight}{\full}$), $\Delta f=$ 5 $\frac{N}{mm.s}$ ($\textcolor{blue}{\full}$), and $\Delta f=$ 2 $\frac{N}{mm.s}$ ($\textcolor{black}{\full}$).}
\label{fig6:time}
\end{figure}

The plate is discretized using 300, 800, and 3500 elements, shown in grey, blue, and black, respectively in figure \ref{fig6:mesh}. The results indicate that mesh convergence is achieved with 800 elements. Moreover, halving the number of time steps per cycle does not lead to a significant change in the force–displacement response, as illustrated in figure \ref{fig6:time}. The staggered algorithm is set to find converge in all subproblems, which provides confidence in the accuracy of the obtained results.

\begin{figure} [htb!]
\centering \includegraphics[scale=0.9]{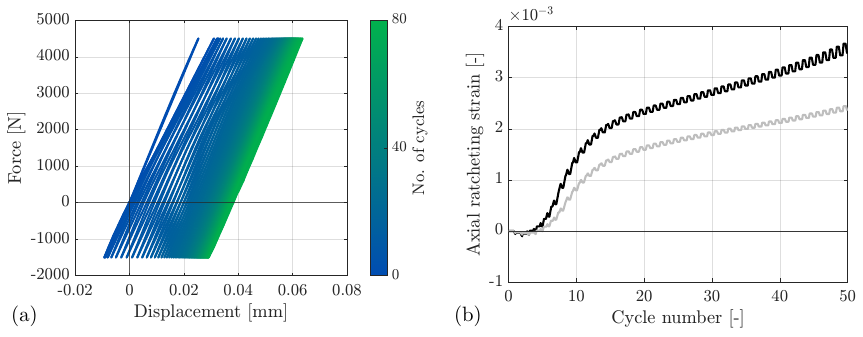}
\caption{(a) Force-displacement response of the perforated specimen under cyclic loading, and (b) evolution of axial ratcheting strain $\Ber_{yy}$ at point A for two parameter sets: $\beta_K=0.2$, $\beta_\mu=0.1$ ($\textcolor{greylight}{\full}$), and $\beta_K=0.3$, $\beta_\mu =0.2$ ($\full$).}
\label{fig6:platefd}
\end{figure}

The force-displacement curve over 78 cycles is presented in figure \ref{fig6:platefd}(a), illustrating ratcheting under force-controlled loading and stiffness degradation due to fatigue. The axial ratcheting strain $\Ber_{yy}$ at point A (indicated in figure \ref{fig6:perf_bc}(a)), where the maximum ratcheting strain occurs, is plotted for two different sets of ratcheting parameters. As expected, higher ratcheting parameters correspond to greater ratcheting strain. This behavior is characteristic of low-cycle fatigue, where inelastic strains accumulate significantly within a relatively small number of loading cycles, leading to progressive material degradation and stiffness loss.  The results demonstrate the model's sensitivity to the parameters $\beta_K$ and $\beta_\mu$, highlighting the importance of their careful calibration when modeling the cyclic plasticity with ratcheting.

\begin{figure} [htb!]
\centering \includegraphics[scale=0.85]{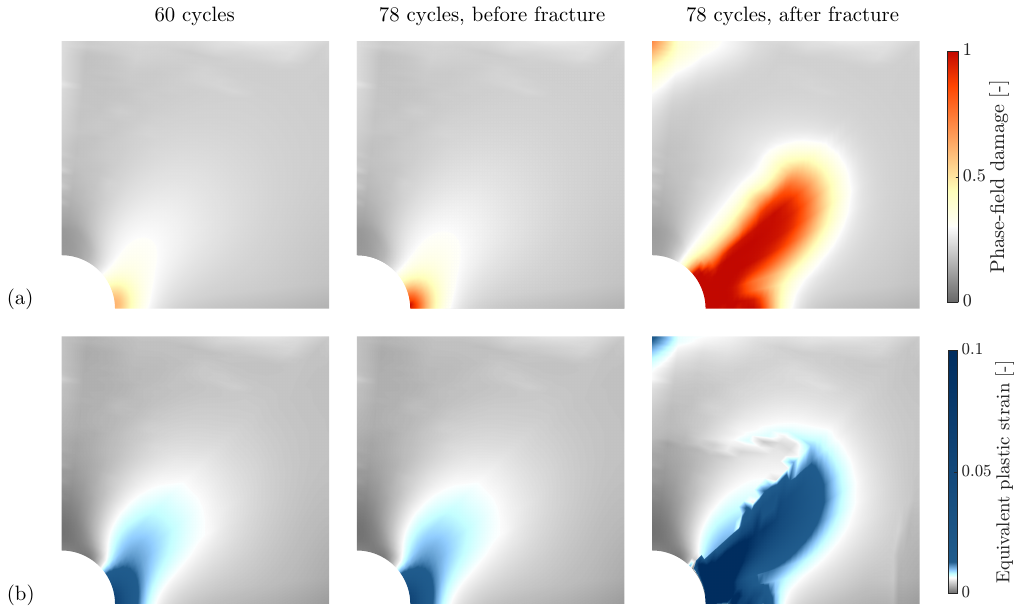}
\caption{Fracture processes in the perforated specimen at different loading cycles, showing (a) damage profiles and (b) equivalent plastic strain distributions.}
\label{fig6:plotplate}
\end{figure}

To comprehensively analyze the cyclic response of the specimen, contours of two parameters are presented in figure \ref{fig6:plotplate} over three different stages of loading to illustrate the fracture process. The first row displays damage profiles and the second row shows distributions of the equivalent plastic strain. As shown in the damage profiles (figure \ref{fig6:plotplate}(a)), failure initiates at stress concentration points around the hole. The equivalent plastic strain contours (figure \ref{fig6:plotplate}(b)) similarly indicate that plastic strains localize in the same region and propagate in an inclined pattern, which is typical for deviatoric plastic flow \cite{ulloa2021b}. This behavior occurs because deviatoric plasticity tends to localize along shear planes, leading to inclined propagation of plastic strains near stress points such as holes.

%

\section{Conclusions}\label{ch6:conc}
This paper presented an extension of the micromechanics-based variational phase-field model of fatigue to incorporate cyclic plasticity with ratcheting in quasi-brittle materials. The ratcheting mechanism was introduced through the evolution of ratcheting strain, which accumulates over loading cycles and captures the inelastic strain growth characteristic of cyclic plasticity. This mechanism is particularly important for low-cycle fatigue, where the accumulation of inelastic strains plays an important role in the progression to final failure. The non-associative ratcheting formulation allows the use of two independent parameters to control the deviatoric and volumetric components of ratcheting, offering additional modeling flexibility.

A series of numerical simulations was conducted to evaluate the model under monotonic and cyclic loading. First, a three-point bending test was simulated to validate the model against experimental data. The same setup was then revisited under cyclic loading to assess the fatigue response. To investigate high-cycle loading, a single element was analyzed and the corresponding $S-N$ curve was obtained. The role of ratcheting was examined through material point simulations under various cyclic loading scenarios. In strain-controlled compression, the model reproduced stress relaxation, while in stress-controlled compression it captured ratcheting behavior. Additional simulations under tension–compression cycles further demonstrated the ability of the framework to simultaneously capture fatigue damage and cyclic plasticity. As a final example, a perforated specimen subjected to force-controlled tension–compression cycles was analyzed, with the model successfully capturing the evolution of localized damage and inelastic strain around stress concentration zones.

\section{Acknowledgments}

This research was funded by Research Foundation - Flanders (FWO project No. G088920N).

\clearpage
\small

\end{document}